\documentclass{nature}

\usepackage{graphicx}
\usepackage{multirow}
\graphicspath{ {images/} }
\DeclareGraphicsExtensions{.png,.pdf,.eps}

\usepackage{hhline}
\usepackage{epstopdf}
\usepackage{amsmath,amssymb}

\usepackage{comment}
\usepackage{url}

\usepackage{xcolor}
\newcommand*{\red}{\textcolor{red}}

\title{\red{Original version submitted in June 2024, see final version at DOI: s41586-024-08480-z.\\
\\}
JWST sighting of decameter main-belt asteroids and view on meteorite sources}

\author{%
Artem~Y.~Burdanov$^{1,*}$, 
Julien de Wit$^{1,*}$, 
Miroslav Brož$^{2}$, 
Thomas G. Müller$^{3}$, 
Tobias Hoffmann$^{4}$, 
Marin Ferrais$^{5}$, 
Marco Micheli$^{6}$, 
Emmanuel Jehin$^{7}$, 
Daniel Parrott$^{8}$, 
Samantha N. Hasler$^{1}$, 
Richard P. Binzel$^{1}$, 
Elsa Ducrot$^{9}$, 
Laura Kreidberg$^{10}$, 
Michaël Gillon$^{11}$, 
Thomas P. Greene$^{12}$, 
Will M. Grundy$^{13}$,  
Theodore Kareta$^{13}$,  
Pierre-Olivier Lagage$^{9}$, 
Nicholas Moskovitz$^{13}$, 
Audrey Thirouin$^{13}$, 
Cristina A. Thomas$^{14}$, and 
Sebastian Zieba$^{10,15}$. 
} 

\begin{document}

\maketitle
\vspace{-3mm}
\textsl{$^*$These authors contributed equally to this work.}
\vspace{-0mm}
\begin{affiliations}

\item Department of Earth, Atmospheric and Planetary Sciences, Massachusetts Institute of Technology, Cambridge, MA, USA;

\item Charles University, Faculty of Mathematics and Physics, Institute of Astronomy, V~Holešovičkách 2, 18000 Prague~8, Czech Republic;

\item Max-Planck-Institut f\"{u}r extraterrestrische Physik, Giessenbachstra{\ss}e, Postfach 1312, 85741 Garching, Germany;

\item Department of Medical Physics and Acoustics, Carl von Ossietzky University of Oldenburg, 26111 Oldenburg, Germany;

\item Florida Space Institute, University of Central Florida, 12354 Research Parkway, Partnership 1 building, Orlando, FL 32828, USA;

\item ESA PDO NEO Coordination Centre, Largo Galileo Galilei, 1, 00044 Frascati (RM), Italy;

\item Space sciences, Technologies \& Astrophysics Research (STAR) Institute University of Liège Allée du 6 Août 19, 4000 Liège, Belgium;

\item Tycho Tracker, Parrott's Studio, LLC;

\item Université Paris-Saclay, Université Paris-Cité, CEA, CNRS, AIM, Gif-sur-Yvette, 91191, France;

\item Max-Planck-Institut für Astronomie, Königstuhl 17, D-69117 Heidelberg, Germany;

\item Astrobiology Research Unit, University of Liège Allée du 6 Août 19, 4000 Liège, Belgium;

\item Space Science and Astrobiology Division, NASA’s Ames Research Center, M.S. 245-6, Moffett Field, 94035, CA, USA

\item Lowell Observatory, 1400 West Mars Hill Road, Flagstaff, AZ 86001, USA;

\item Northern Arizona University, Department of Astronomy \& Planetary Science, PO Box 6010, Flagstaff, AZ 86011, USA;

\item Leiden Observatory, Leiden University, Niels Bohrweg 2, 2333CA Leiden, The Netherlands.

\end{affiliations}
\vspace{-0mm}
\begin{abstract}

Asteroid discoveries are essential for planetary-defense efforts aiming to prevent impacts with Earth\cite{Cheng_2018P&SS..157..104C},
including the more frequent\cite{Brown2002} megaton explosions
from decameter impactors\cite{Chyba_1993Natur.361...40C,Chao_1960Sci...132..220C,Stoffler_2002M&PS...37.1893S,Reddy2019}.
While large asteroids ($\geq$100\,km) have remained in the main belt since their formation\cite{Bottke_2005Icar..175..111B}, small asteroids 
are commonly transported to the near-Earth object (NEO) population\cite{Farinella_1998Icar..132..378F,Chesley_2003Sci...302.1739C}.
However, due to the lack of direct observational constraints,
their size-frequency distribution
---which informs our understanding of the NEOs and the delivery of meteorite samples to Earth---
varies significantly among models\cite{Brown_2013Natur.503..238B,Harris_2021Icar..36514452H,Nesvorny_2024Icar..41115922N,Broz_2023,Marsset_2023}.
Here, we report 139 detections of the smallest asteroids ($\gtrapprox $10\,m)
ever observed in the main belt, which were enabled by JWST's infrared capabilities
covering the asteroids' emission peaks\cite{Mueller2023A&A}
and synthetic tracking techniques\cite{tyson1992limits,shao2014finding,Burdanov2023MNRAS}.
Their size-frequency distribution exhibits a break at ${\sim}100\,{\rm m}$
(the debiased cumulative slopes
$q = -2.25$ and $-0.98$, with the uncertainty ${\sim}0.1$),
suggestive of a population driven by collisional cascade.
These asteroids were sampled from multiple asteroid families
---most likely Nysa, Polana and Massalia---
according to the geometry of pointings considered here.
Through additional long-stare infrared observations,
JWST is poised to serendipitously detect thousands of decameter-scale asteroids across the sky,
probing individual asteroid families\cite{Nesvorny_2015aste.book..297N}
and the source regions of meteorites\cite{Broz_2023,Marsset_2023} ``in-situ''.

\end{abstract}

\flushbottom
\maketitle



\thispagestyle{empty}

\begin{figure}[!ht]
    \centering
    \hspace*{0mm}\includegraphics[scale=0.58]{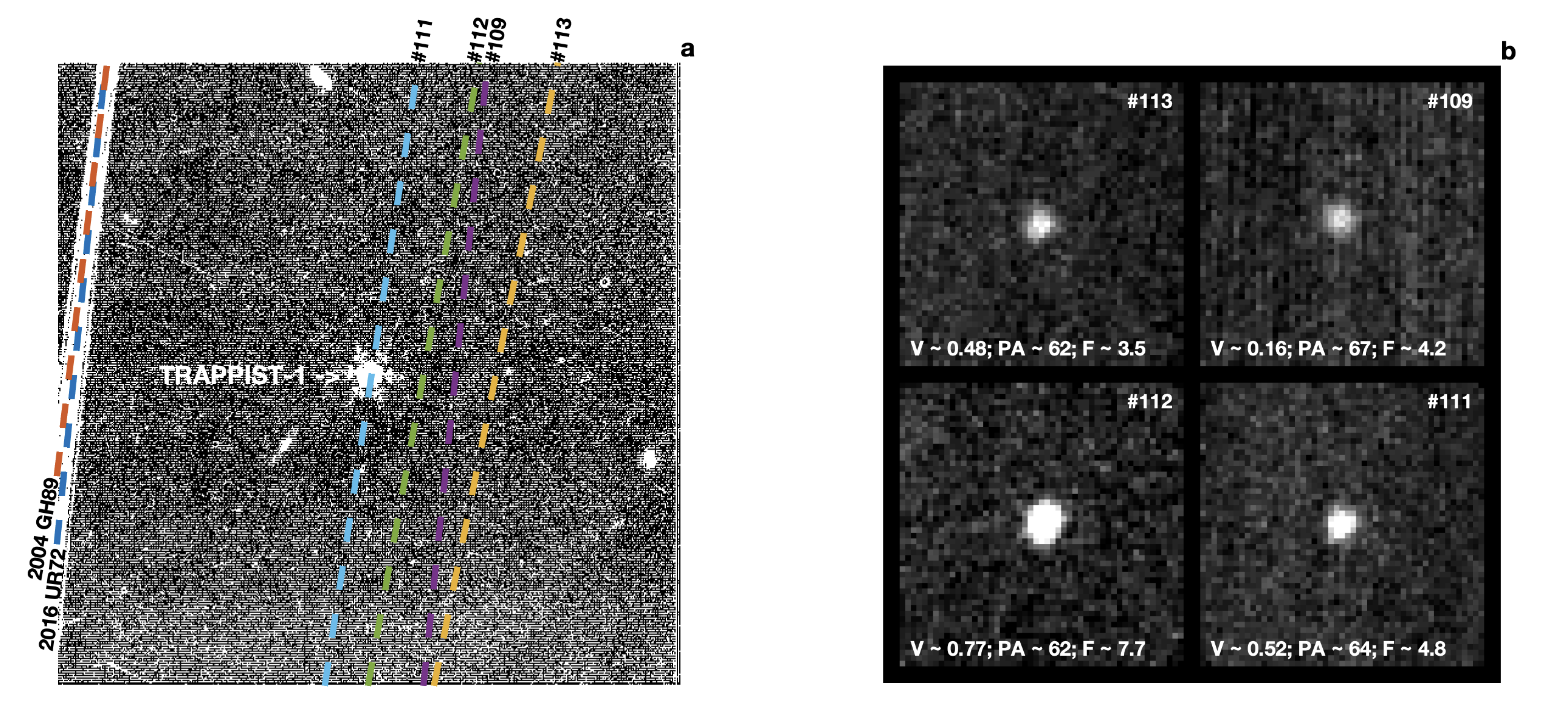}
    \vspace*{0mm}\caption{
    {\bf Basics of a blind search for asteroids using synthetic tracking.} \textbf{a.} Average stack of exposures 4,000 to 4,500 from PID 3077 centered on the ultra-cool star TRAPPIST-1, revealing two known bright asteroids (2004 GH89 and 2016 UR72) crossing the left side of the field of view (FoV). Being bright, they are detectable on individual exposures, leading to a trail on the stacked exposure marked by the orange and blue dotted lines. The other dashed lines refer to the paths of four unknown asteroids crossing the FoV at the same time, but only detectable in stacked exposures that are first shifted along their respective paths, which are identified via a blind search through the ``shift-and-stack'' technique. \textbf{b.} Shifted-and-stacked exposures centered on four new asteroids (\#113, \#109, \#112, and \#111) with their speed (V, in arcsec/min), position angle (PA, in degree), and Flux (F, in $\mu$Jy). All the properties of the 139 new asteroids are reported in Supplementary Table 1.
    }
    \label{fig:fig01}
\end{figure}

Asteroids are discovered by their motion relative to the background stars. This observed motion results from asteroids' actual orbital movement combined with motion induced by Earth's (and/or a satellite's) parallactic movement. While most asteroid-search projects detect objects in single images (exposures) and link their motion across multiple images, this method may miss fainter objects which are not visible on an individual image. To address this, the "shift-and-stack" technique, developed in the 1990s, enhances the signal-to-noise ratio (SNR) by combining multiple images into one "stack" image\cite{tyson1992limits,Gladman1997A&A,Bernstein2004AJ}. This method involves predicting the asteroid's motion, shifting image pixels accordingly, and then combining the images (Fig.~\ref{fig:fig01}). Synthetic tracking, an extension of ``shift-and-stack" technique, does not rely on prior knowledge of an asteroid's motion, but rather performs a fully ``blind" search by testing a series of possible shifts\cite{shao2014finding,Zhai2014ApJ,Heinze2015AJ} (i.e., velocity vectors). However, this method's computational intensity posed a bottleneck until the widespread availability of graphics processing units (GPUs). The subsequent usage of GPU-based synthetic tracking increases the scientific return of monitory campaigns, such as exoplanet transit-search surveys, by recovering serendipitous asteroid detections\cite{Burdanov2023MNRAS,2023MNRAS.526.3601H}.

\begin{figure}[!ht]
    \centering
    \hspace*{0mm}\includegraphics[scale=0.5]{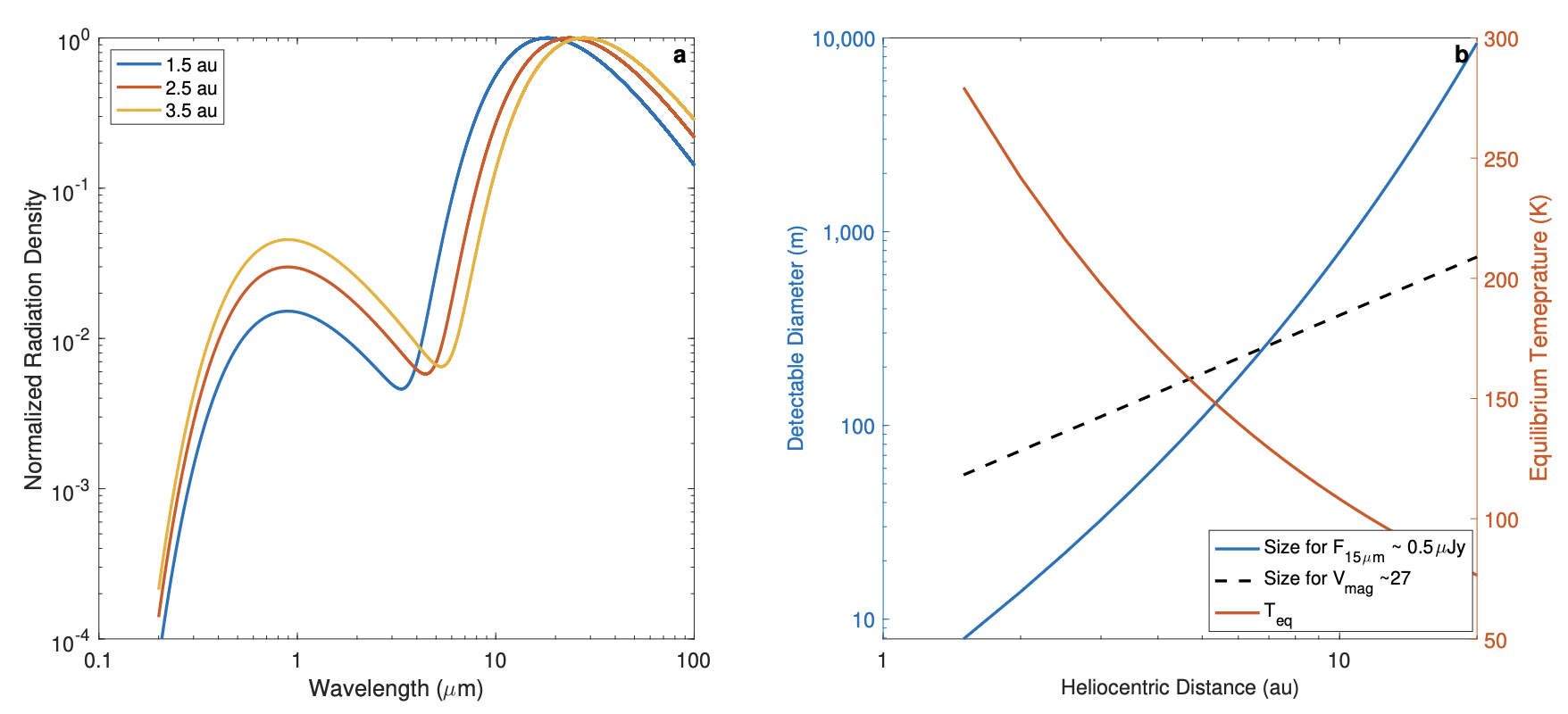}
    \caption{\textbf{JWST's far-infrared window into the main-belt asteroid population.} \textbf{a.} Radiation Density (Jy) normalized to the peak emission for 0.1-albedo asteroids with an heliocentric distance of 1.5 (blue), 2.5 (red), and 3.5\,au (yellow) showcasing the favorable infrared-to-visible flux ratio. \textbf{b.} Minimum radius of an asteroid detectable for a 0.5\,$\mu$Jy detection threshold at 15\,$\mu$m compared to state-of-the-art capabilities in the visible---dotted line represents the radius detection threshold at Vmag$\sim$27 (Ref.\cite{Reddy2019}). With a 0.5\,$\mu$Jy detection threshold (Extended Data Figure\,\ref{fig:recovery-rate}), JWST can outperform searches in the visible up to 10 au, and by up to two orders of magnitude in size in the main belt.  
    }
    \label{fig:fig02}
\end{figure}

The vast majority of known asteroids have been discovered by ground-based surveys at visible wavelengths. Asteroids' full spectral energy distributions are a combination of reflected sunlight (driven by the object's albedo) and thermal emission, with the central wavelengths of the thermal peak ranging between 5 and 20\,$\mu$m for objects between 1 and 10\,au  (Fig.~\ref{fig:fig02}). With an exquisite sensitivity in that wavelength range and a large aperture, JWST is ideal for detecting the thermal emission of asteroids and revealing the smallest main-belt asteroids (MBAs)\cite{Mueller2023A&A}. Such observations combined with orbital information can yield accurate radius estimates, which are less affected
 by degeneracy with the albedo than those from visible-light
 observations. Indeed, the visible-light detection of a typical MBA with known orbit can
 be explained by an object with a small size with high albedo or a large size
 with low albedo. For the wide range of albedos from 3-40\%, the corresponding
 sizes can vary by a factor 3-4. In contrast, a thermal infrared (IR) measurement close to the object's thermal peak constrains the object's size to within about 10-20\% (see Methods).

JWST observing programs conducted with no dithering are especially suitable for synthetic tracking as all exposures from one visit can be shifted and stacked. This makes JWST sensitive to small IR fluxes from moving objects in a field of view (FoV) and enhances its capability to detect faint asteroids. Such a dithering-free long-stare mode was used to observe the TRAPPIST-1 star (located 0.6\,deg from the ecliptic) with the MIRI instrument\cite{Rieke2015} at 15\,$\mu$m as part of multiple programs aimed at characterizing the TRAPPIST-1 exoplanetary system through measurements of the inner planets' dayside emission (Program IDs 1177 and 2304, with Greene, and Kreidberg as PI, respectively) and their combined thermal phase curve (PID 3077, PI Gillon).
In total, JWST observed the TRAPPIST-1 star for 93.5\,hours during 11 visits in 2022-2023. After applying our GPU-based framework for detecting asteroids in targeted exoplanet surveys\cite{Burdanov2023MNRAS,2023MNRAS.526.3601H}, we were able to detect 8 known and 139 unknown asteroids which happened to serendipitously cross MIRI FoV of $56.3'' \times56.3''$ or $112'' \times113''$ (depending on the particular observing program). The known objects are MBAs with fluxes between 100 and 1,700 $\mu$Jy and diameters (D) between 200 and 2,500\,m (Extended Data Table~\ref{tbl:8ast}).


The 139 new detections could not be attributed to any known asteroids, where we searched for previously discovered objects positioned closer than 1' from each detection (see Methods). IR fluxes of these new objects range from $0.5\,\mu$Jy to $600\,\mu$Jy, with a detection/sensitivity threshold at ${\sim}0.5\,\mu$Jy (see Methods and Extended Data Figure\,\ref{fig:recovery-rate}).
Our detections spend from 30\,min to 8\,hours in the MIRI FoV. Even in the case of the longest observing arc of 8\,hours, orbits of different dynamical classes can fit the data well and are statistically indistinguishable. We thus used ensembles of known objects which were predicted to be in a $6^{\circ}\times2^{\circ}$ area around the TRAPPIST-1 star at the time of detection of an unknown asteroid as priors to derive posterior probability distributions on the distance from JWST to each unknown asteroid. This methodology yielded the distance with a typical uncertainty of 0.2-0.3\,au and adequately returned the distance of the eight known asteroids (see Methods). We estimated distances from JWST to unknown asteroids to be from 0.9 to 3.0\,au placing them in the main asteroid belt, with diameters ranging from 10 to 500\,m (see Fig.~3 and Methods). The detection/sensitivity threshold at $\sim$0.5\,$\mu$Jy thus translates into an observational bias emerging in the 20- to 40-m diameter regime with a sharp cutoff by $\sim$10\,m.


The size-frequency distribution (SFD) of our asteroid detections is unusually shallow
at sizes larger than 100\,m,
corresponding to a population depleted by collisions (Fig.~\ref{sfd_0200.000_jwst}).
It can be described by a power law,
$N({>}D) = CD^q$,
with the exponent
$q = -0.98 \pm 0.14$.
On the other hand, the observed SFD is significantly steeper below 100\,m,
with $q \simeq -1.45$.
Taking into account non-negligible uncertainties of individual diameters
(see Methods),
we also computed a debiased exponent
$q = -2.25 \pm 0.07$,
valid between 100\,m and approximately 10\,m.

This steeper slope is characteristic of the strength regime of fragmentation\cite{Dohnanyi_1969JGR....74.2531D}.
In fact, bodies ${\sim}100\,{\rm m}$ in size are among the weakest
in the Solar System\cite{Benz_1999Icar..142....5B}.
Their studies thus provide unique insights into realistic asteroidal materials.

At decameter sizes, small asteroids are most likely related to recent disruptions and known asteroid families\cite{Nesvorny_2015aste.book..297N,Broz_2023}.
The associations are based on the same methodology;
on orbits of known objects located close to the JWST field of view
(Fig.~\ref{all_ai_nature}).
More specifically,
the Nysa family was sampled,
together with
Polana,
Massalia,
Koronis2, and/or
Karin
(Supplementary Table 1).
In particular, the synthetic size-frequency distribution (SFD) of the Nysa and Polana families exhibit
similar slopes that match the observations in both the shallow and steep regimes
(Fig.~\ref{sfd_families}).
According to the collisional model from ref.~\cite{Broz_2024},
their ages are of the order of 200 and 600\,My, respectively.
Consequently, sub-km bodies should be in a collisional equilibrium.
Other families exhibit a variety of slopes,
especially because some of them are young\cite{Broz_2023,Marsset_2023}.
At the decameters sizes, we thus expect the Massalia family is dominant.
Nonetheless, when asteroids are sampled from multiple families,
the resulting SFD is indeed a combination of steep and shallow slopes,
resembling the distribution of the whole asteroid belt
(Fig.~\ref{sfd_0200.000_jwst}).

Compared to previous pencil-beam surveys%
\cite{Gladman_2009Icar..202..104G,Ryan_2015A&A...578A..42R,Garcia_2024A&A...683A.122G},
our observations reveal on one hand a continuation of the shallow slope
down to much smaller sizes than previously thought.
For example, observations obtained by the Hubble Space Telescope\cite{Garcia_2024A&A...683A.122G}
were only complete to V$\sim$ 23\,${\rm mag}$,
corresponding to approximately $500\,{\rm m}$, and reaching as far as V$\sim$28\,${\rm mag}$---in comparison to V$\sim$27\,${\rm mag}$ for ground-based surveys\cite{Reddy2019} (Fig.\,3.b).
On the other hand, when debiasing the JWST observations,
we find the slope is steeper than previously thought, with a clear slope increase around 100\,m.
It is suggestive of the first and long-awaited evidence of a population,
which is evolving by collisions and at the same time moving to the NEO space.\cite{Nesvorny_2024Icar..41115922N}




Looking ahead, it is anticipated that JWST will be observing 15-20 exoplanet host stars for at least 500\,hours with MIRI\cite{Redfield2024}, following a similar observing strategy to the one followed to acquire the data used here\cite{Greene_2023Natur.618...39G,Zieba_2023Natur.620..746Z}. As a large fraction of the host stars amenable for such studies are fortuitously found within 20 degrees of the ecliptic, these 500\,hours will yield hundreds of additional decameter asteroids. In addition, an average of $\sim$1800\,hours of MIRI observations were gathered per Cycle in Cycles 1 through 3. We thus expect that
JWST will detect thousands of decameter asteroids and---with multi-visit observation strategy making it possible to recover orbits precisely---will
probe different families
and study the source regions of meteorites\cite{Broz_2023,Marsset_2023,Broz_2024} ``in-situ''.

Beyond the detection of asteroids otherwise undetectable, JWST can also yield exquisite infrared rotation curves of large (i.e., $\gtrapprox$300\,m) asteroids thereby allowing their further characterization (see Extended Data Figure\,\ref{lc_YP90}). The insights gained from JWST's infrared rotation curves of asteroids will be particularly valuable in two ways. First, their precision vastly exceeds their counterpart in the visible. Second, unlike visible rotation curves they are mostly insensitive to surface topology---which results in reduced degeneracy with the albedo (similarly to the size-estimation process in the infrared). As rotation rates can inform the origin and evolution of asteroid family\cite{Farinella1992,Bottke2006,Carruba2020}, their relationship to the break-up barrier\cite{Polishook2014}, and even be used to derive their internal properties during close encounters\cite{Dinsmore2023} JWST is also poised to further our understanding of asteroid families and meteorite source bodies through rotation measurements. 

There is thus a great deal of synergy between JWST and its infrared capabilities and other facilities dedicated to the study of minor bodies in the visible, such as the Rubin observatory\cite{LSST_2009arXiv0912.0201L},
an all-sky survey aiming at discoveries of 100-meter asteroids.
Combining these expected discoveries of such dedicated facilities with the capabilities of JWST presented here will finally allow to constrain dynamical and collisional models all the way down to 10\,m. 

This synergy will extend beyond scientific endeavors and support planetary-defense efforts. Although planetary-defense often appears associated with preventing events such as the impact that led to the extinction of the dinosaurs\cite{Alvarez1980}, decameter objects offer non-negligible threats which occur at much higher rates---rates are proportional to $D^{-2.7}$, leading to decameter impactors being $\sim$10,000 times more frequent than km-sized ones\cite{Brown2002}. Decameter objects can lead to megaton explosions leaving behind km-sized craters. They are in fact at the origin of recent events of importance on Earth like
Chelyabinsk\cite{Brown_2013Natur.503..238B},
Tunguska\cite{Chyba_1993Natur.361...40C},
Barringer\cite{Chao_1960Sci...132..220C}, or
Steinheim\cite{Stoffler_2002M&PS...37.1893S}. JWST's capability to detect---and thus monitor---decameter objects all the way to the main belt and meteorite source bodies will hence be a pivotal asset for planetary-defense efforts.






\clearpage

\section{Main Figures}

\clearpage

\clearpage

\begin{figure}[!ht]
    \centering
    \vspace*{-10mm}\hspace*{-25mm}\includegraphics[scale=.5]{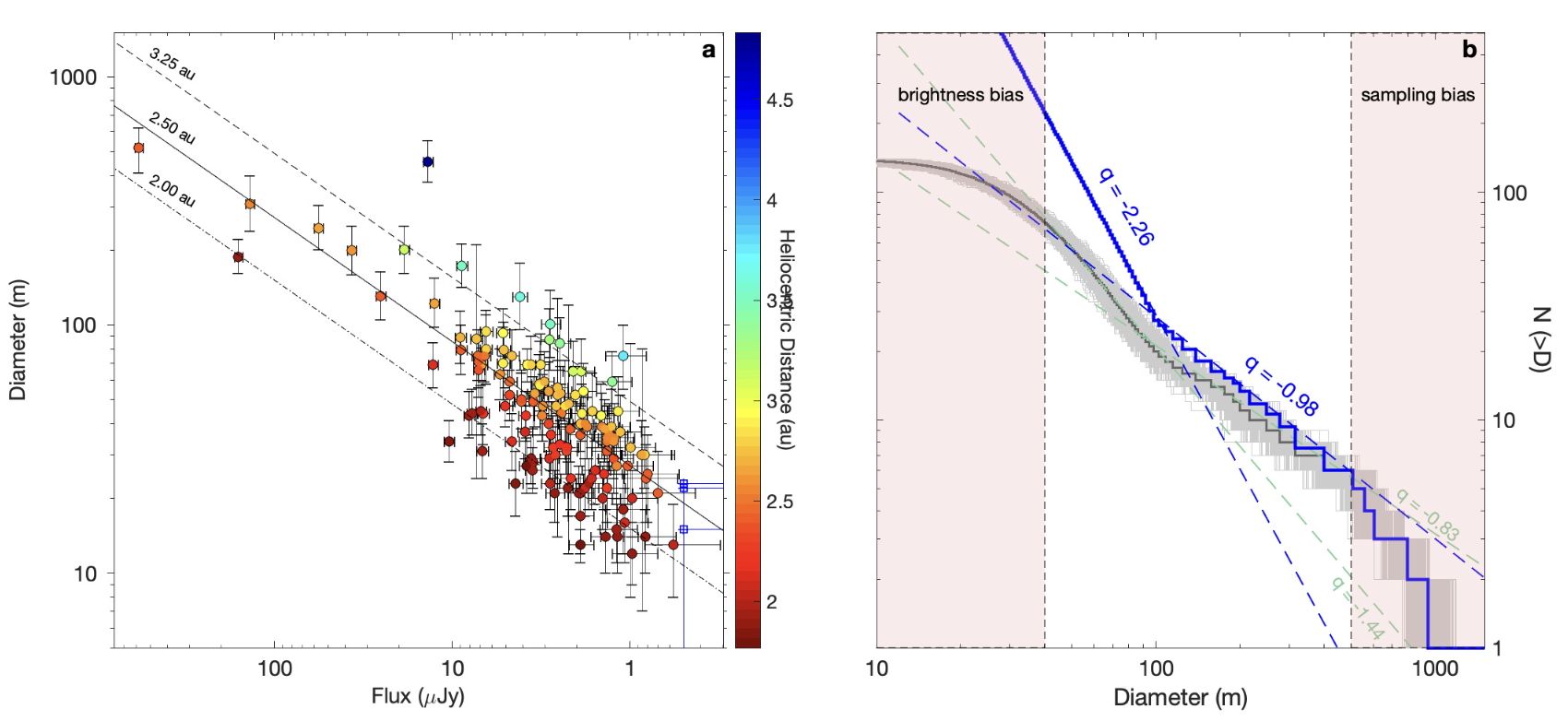}
    \caption{
    \textbf{Flux-diameter and size-frequency relationships for the 139 new asteroids.}
    \textbf{a.} Fluxes, diameters, and heliocentric distances of the new asteroids range from 0.5 to 600\,$\mu$Jy, 10 to 500\,m, and 1.8 to 4.5\,au, respectively. The dash-dot, solid, and dotted lines represent the size-flux relationships for objects at 2.00, 2.50, and 3.25\,au, respectively. The large uncertainty on the size of bright asteroids despite precise flux measurements stem from a large uncertainty associated with the orbital configuration of these new objects. Detections beyond the sensitivity threshold (${\sim}\,0.5\,\mu$Jy, see Extended Data Figure\,\ref{fig:recovery-rate}) are reported in blue as upper limits on the flux and radius. \textbf{b.} Ensemble of cumulative size-frequency distributions (SFDs) built from 1,000 1-$\sigma$-perturbed asteroid diameters in order to propagate the size uncertainties onto the SFD estimate (median in grey) via the Monte Carlo method\cite{Metropolis1949} together with the debiased SFD (blue)---see Methods for details. The SFD presents two distinct regimes with exponents $q = -2.25 \pm 0.14$ and $q = -0.98 \pm 0.07$ (dashed lines)---$N({>}D) = C D^q$.
    }
    \label{fig:fig03}
\end{figure}

\clearpage


\clearpage

\definecolor{aquamarine}{RGB}{0,255,204}
\definecolor{lime}{RGB}{128,255,0}

\begin{figure}
\centering
\includegraphics[width=12cm]{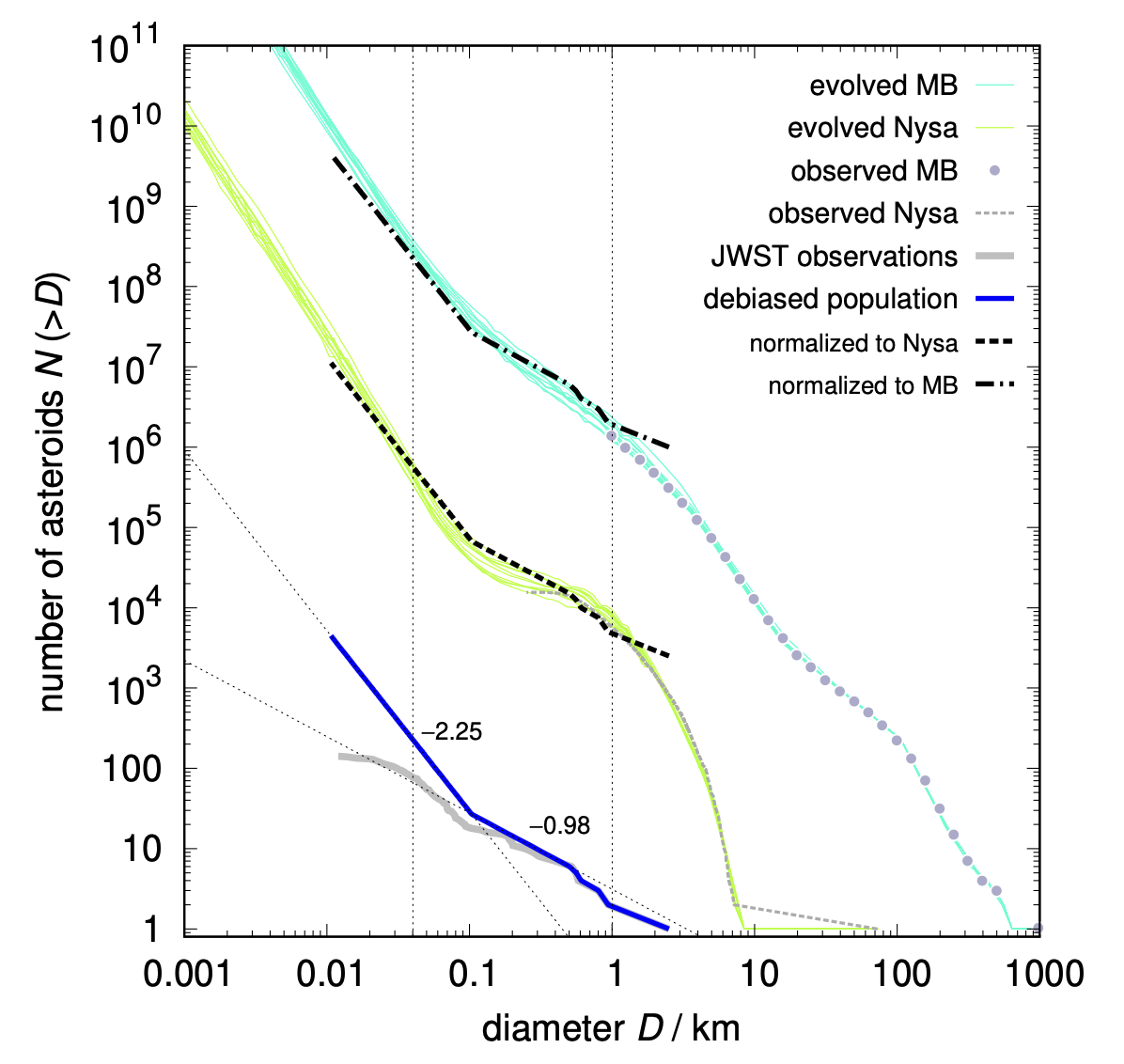}
\caption{
{\bf Asteroids observed by the JWST have a size-frequency distribution
with a break at ${\sim}100\,{\rm m}$,
revealing a population in collisional equilibrium.}
The observed (gray) and debiased (\textcolor{blue}{blue}) cumulative SFDs,
$N({>}D) = C D^q$,
are shown together with corresponding slopes~$q$ (dotted).
According to the collisional model from ref. \cite{Broz_2023},
this exactly corresponds to the main belt population
between 1,000 and approximately 50\,m
(\textcolor{aquamarine}{aquamarine}).
Additionally, evolved asteroid families exhibit similar exponents
due to ongoing collisions with the main belt population.
The Nysa family is plotted for comparison
(\textcolor{lime}{lime}),
with a different normalization of the JWST data
(dashed).
}
\label{sfd_0200.000_jwst}
\end{figure}

\clearpage

\begin{addendum}
\item[Acknowledgements] This work is based in part on observations made with the NASA/ESA/CSA JWST. The data were obtained from the Mikulski Archive for Space Telescopes (MAST) at the Space Telescope Science Institute (STScI), which is operated by the Association of Universities for Research in Astronomy, Inc., under NASA contract NAS 5-03127 for JWST. These observations are associated with programs 1177, 2304, and 3077 with
Greene, Kreidberg, and Gillon as PI, respectively. 
A.Y.B. and J.d.W. thanks Michael~J.~Person for discussions regarding astrometry and Scott Stuart for initial SFD calculations. J.d.W. and MIT gratefully acknowledge financial support from the Heising-Simons Foundation, Dr. and Mrs. Colin Masson and Dr. Peter A. Gilman for Artemis, the first telescope of the SPECULOOS network situated in Tenerife, Spain. This work has been supported by the Czech Science Foundation through grant 21-11058S (M. Brož). This work has been supported by the NVIDIA Academic Hardware Grant Program. TRAPPIST is funded by the Belgian National Fund for Scientific Research (F.R.S.-FNRS) under grant PDR T.0120.21, and the university of Liège. E.J. is a Belgian FNRS Senior Research Associate. T.H. gratefully acknowledges financial and technical support from ESA, the Erasmus+ program, co-funded by the European Union, and the Cusanuswerk Bischöfliche Studienförderung, funded by the German Federal Ministry of Education and Research.

\item[Author Contributions] A.Y.B. and J.d.W. designed and led the study. A.Y.B. performed the data reduction and asteroid detection (incl., injection-retrieval tests) following Ref.\cite{Burdanov2023MNRAS} with support from J.d.W.. J.d.W. designed the population-based distance estimation with support from A.Y.B., provided the preliminary size estimations, and performed the information-content analysis for the size-frequency distribution with support from M.B. M.B. performed the interpretation of the size-frequency distribution. T.G.M. derived the final size and albedo estimates. T.H. provided the debiased H magnitudes. EJ put in relation the exoplanet team with asteroids experts and led the ground based follow-up observations together with A.Y.B., M.F., W.M.G., T.K., N.M., A.T., and C.A.T. M.F. led the rotation curve analysis. M.M. provided the detailed orbital estimations. D.P. and S.N.H. provided complementary target identifications, flux estimates, and support for Tycho Tracker. All authors contributed to the writing of the manuscript. 

\item[Competing Interests]
The authors declare that they have no competing financial interests.

\item[Data Availability]

\item[Code Availability]
This work makes use of the following publicly available codes: 
NumPy\cite{harris2020array}, Matplotlib\cite{Hunter:2007}, Astropy\cite{2013A&A...558A..33A, 2018AJ....156..123A}, SciPy\cite{2020SciPy-NMeth}, Pandas\cite{reback2020pandas,mckinney-proc-scipy-2010}, Astroquery\cite{2019AJ....157...98G}.

\item[Correspondence]
Correspondence and requests for materials should be addressed to jdewit@mit.edu. 

\end{addendum}

\newpage
\pagebreak
\clearpage

\begin{methods}

\subsection{JWST image processing}

Data from Program IDs (PIDs) 1177 and 2304 were acquired with JWST/MIRI in October and November 2022 using FULL subarray mode resulting in $112'' \times113''$ (1024$\times$1032 pixels$^2$) FoV. Data from PID 3077 were obtained in November 2023 in BRIGHTSKY subarray mode resulting in a smaller FoV of $56.3''\times56.3''$ (512$\times$512 pixels$^2$). All programs used F1500W filter and FASTR1 readout mode. 

For our asteroid search, we downloaded exposure raw data products ("*uncal.fits” files) from the Barbara~A.~Mikulski Archive for Space Telescopes. Then, we ran Stage 1 and 2 JWST Science Calibration Pipeline version 1.13.4 to produce calibrated "*calints.fits" files. These files are single exposures containing results for all integrations in an exposure with world coordinates and photometric information. Every "*calints.fits" 3-D data file was sliced into a set of 2-D data files containing the pixel values for each integration. For PIDs 1177 and 2304, we trimmed 2-D data files to exclude parts of the detector designed for coronagraphic imaging resulting in a useful FoV of $72''\times113''$ (654$\times$1032~pixels$^2$). Each 2-D data file was then corrected for sky background using Photutils\cite{larry_bradley_2023_7946442} Python software package. Background-subtracted images were then searched for any moving objects which happen to cross the FoV.



\subsection{Detection of asteroids using synthetic tracking and their flux estimation}

We leveraged our custom-build wrapper\cite{Burdanov2023MNRAS,2023MNRAS.526.3601H} around the Tycho Tracker\cite{parrott2020tycho} synthetic tracking software to explore a wide range of motion vectors and generating trial exposure stacks for each vector. No restrictions on position angle (PA) were used and speed was in a range of 0.001-1.2 arcsec/min. Each trial exposure stack was computed by shifting the exposures according to the motion offsets associated with the current vector. If an object had motion similar to that of the given vector, it was extracted from the trial exposure stack by the detection process. We grouped images to detect faster moving objects by dividing the image sequence from one JWST visit into overlapping groups of 100 exposures, ensuring the detection of an asteroid if it appears in at least 50\% of the group's images. After detecting fast-moving objects, we performed a search for slower moving objects by searching all the images from one JWST visit (up to 350 exposures for PIDs~1177 and 2304, and up to 1000 exposures for PID~3077). A set of candidate detections (tracks) was returned with corresponding speed, PA, pixel coordinates, and SNR of detection. 

We cross-matched every track with already known objects using Tycho Tracker capability to use the Minor Planet Center (MPC) database of small body orbits and Find\_Orb\footnote{\url{https://www.projectpluto.com/find_orb.htm}} to predict positions of known object within the FoV. A match was made if a known object was positioned within 0.1\,$'$. We double-checked our matches using the NASA Jet Propulsion Laboratory (JPL) Small-Body Identification Application Program Interface (API)\footnote{\url{https://ssd-api.jpl.nasa.gov/doc/sb_ident.html}}. We were able to detect 8 known asteroids  and 139 unknown asteroids with SNR$\geq5$ (see Table~\ref{tbl:8ast} and Supplementary Table~1).  

We measured the flux of each asteroid using aperture photometry, employing a circular aperture with a diameter of 2.5$\times$FWHM (Full Width at Half Maximum) of the asteroids' point spread function (PSF) in the shifted and stacked image. The sky background was measured in an annulus beyond the asteroid aperture using a median sky fitting algorithm. The annulus had a radius of 4$\times$FWHM and a width of 2$\times$FWHM. We applied similar aperture photometry to estimate the flux of the TRAPPIST-1 star (the only stellar object in the FoV) and calculated the flux ratio of each asteroid to TRAPPIST-1. Previous studies\cite{Greene2023Nature} have shown that the absolute flux of TRAPPIST-1 in the F1500W band is stable at 2,590$\pm$80\,$\mu$Jy. We determined the absolute flux values of the asteroids using this reference flux of TRAPPIST-1. Flux estimates and their associated errors can be found in Table~\ref{tbl:8ast} and Supplementary Table~1.


\subsection{Asteroid detection efficiency}

We conducted a series of injection-recovery tests to evaluate our asteroid detection efficiency. We injected a 5$\times$5 grid of synthetic moving objects, each with various flux values (see below), into 200 sky-subtracted $72''\times113''$ (654$\times$1032~pixels$^2$) FITS files. Synthetic objects were placed in such a way that they spend all the time in FoV, had random PAs from a uniform distribution between 70 and 80\,deg and random speeds from a uniform distribution between 0.2 and 0.3\,arcsec/min (representing a typical speed and PA of real detected unknown asteroids). After running the synthetic tracker, we compared the detected objects with the injected ones. We repeated this test four more times for the same flux value as a sensitivity analysis for our estimated recovery rate as a function of flux. In total, we performed six sets of injection-recovery tests with objects having flux values of 3.0, 2.0, 1.5, 1.0, 0.5 and $0.11\,\mu{\rm{Jy}}$. Our recovery rate is $100\%$ for objects down to $1.5\,\mu{\rm{Jy}}$ (see~Fig.~\ref{fig:recovery-rate}), which then sharply drops to $50\pm2\%$ at $0.5\,\mu{\rm{Jy}}$. The derived cutoff is 0.5$\,\mu$Jy with an observation bias starting at $\sim 1\,\mu$Jy (12 out of 139 unknown objects have fluxes smaller than $1\,\mu{\rm{Jy}}$).


\subsection{Orbit estimations}

To estimate the sizes of the unknown asteroids from their IR fluxes, we require Observer-Target (O-T) distance, i.e. a distance from the JWST to an unknown asteroid at the time of observations. Due to the short duration of the asteroids' arcs, a large ensemble of possible orbital configurations exist for each object---even in the case of the longest observing arc of 8\,hours of unknown asteroid \#91. To overcome this bottleneck, we developed a method using distances to ensembles of known objects as priors to derive posterior probability distributions for unknown asteroid distances.

For every JWST visit, we obtained a list of all known asteroids predicted to be within a $6\times2$\,degree$^2$ area (RA$\times$Dec) around the TRAPPIST-1 star using the JPL Small-Body Identification API. We queried JPL Horizons\footnote{\url{https://ssd-api.jpl.nasa.gov/doc/horizons.html}} for the speed, PA, O-T distance,  Sun-Target (S-T) distance, and the corresponding Sun-Target-Object (S-T-O) phase angles for each known asteroid. Then, every unknown asteroid was placed in speed/PA parameter space, and known objects within three different ellipses surrounding the unknown asteroid position in that parameter space were selected (Fg.\,\ref{fig:distance_estimation}): 1-$\sigma$ ellipse is defined as an ellipse with widths of 5\% of the unknown object's speed and an absolute value 0.5\,deg for PA (which correspond respectively to the typical 1-$\sigma$ uncertainties on speed and PA), $3\sigma$ and $10\sigma$ ellipses are $3\times$1-$\sigma$ and $10\times$1-$\sigma$ respectively. We assigned a distance to a particular unknown asteroid as a median value of O-T distances to known asteroids within different ellipses. We did similar calculations to S-T distance, and the corresponding S-T-O phase angles. 

We assessed the sensitivity of our distance estimates to different samples of known asteroids and confirmed negligible dependencies when compared to the derived uncertainties (i.e., spread in O-T distances). As a proof of concept, we tested this method on eight known asteroids. We removed their true speed and PA values from the speed/PA parameter space and treated them as unknown objects. We found that in all instances the true values are retrieved within the derived 1-$\sigma$ error bars (typically between 0.2-0.3\,au). See~Fig.~\ref{fig:distance_estimation} for an application of the method to known asteroid 2004~GH89, which was observed in PID~3077. All the properties of the 139 new detections are reported in Supplementary Table 1.


\subsection{Debiasing asteroids' absolute magnitudes}

The size calculation of known asteroids relies on refined and debiased asteroids' absolute H-mag estimates, which were derived through a novel correction method \textit{DePhOCUS}\cite{Hoffmann2024}. 
The method performs debiasing of astro-photometric observations from the MPC with corrections using a statistical analysis based on an accurate reference of 468 asteroids with more than 450,000 observations in total. The method allows a derivation of 17 updated significant color bands, 90 catalog and 701 observatory corrections (significance level $p=0.90$), which lead to a reduction of more than $36\%$ in the root mean square (RMS) of the asteroids' phase curve and a more accurate estimation of the parameters of the H-G phase curve model, where G is the slope parameter. We used the corrections to debias the observations at the MPC for all known asteroids in the present study and compute the absolute H magnitude values and their uncertainties, assuming $\mathrm{G=0.15}$.


\subsection{Size and albedo determination}

The radiometric analysis of all detected asteroids was done via the Near-Earth Thermal Model (NEATM\cite{1998Icar..131..291H}).
NEATM was originally developed for near-Earth asteroids, but is now
also widely applied to asteroids in the main-belt and beyond (see, e.g. Ref.\cite{Masiero_2011ApJ...741...68M,2017A&A...603A..55A,2018A&A...612A..85A}). 
In this model, the asteroids are approximated by non-rotating and smooth spheres which are in instantaneous thermal equilibrium with the incident solar radiation. This allows to calculate the temperature of each surface
element via $\mu\times(1-A)\times S_{\rm sun}/r^2 = \epsilon\eta\sigma T^4$, 
where $\mu$ is the cosine of the angle between the element’s normal and the direction towards the sun, $A$ the bolometric Bond albedo,
$S_{sun}$ the solar incident energy at 1 au, $r$ the heliocentric distance,
$\epsilon$ the emissivity (a fixed value of $\epsilon$=0.9 is taken), and $\sigma$ the
Stefan-Boltzmann constant. The infrared beaming parameter $\eta$ was
introduced as a free parameter. It can be determined from a fit to
multi-band infrared measurements (as originally done in Ref.\cite{1998Icar..131..291H}),
or calculated from published linear phase-angle relations (e.g., Ref.\cite{2008Icar..193..535W,Masiero_2011ApJ...741...68M,2018A&A...612A..85A}).
For specific asteroid groups average $\eta$ values are often taken, e.g., $\eta$=1.4 for
near-Earth asteroids\cite{2011ApJ...743..156M},  $\eta$=1.2 for
Mars-crossing asteroids\cite{2017A&A...603A..55A},  $\eta$=1.0 for MBAs\cite{Masiero_2011ApJ...741...68M}, $\eta$=0.77 for Hildas and Jupiter Trojans\cite{2012ApJ...744..197G}, or $\eta$=1.2 for trans-Neptunian objects\cite{2012A&A...541A..94V}. Ref.\cite{Masiero_2011ApJ...741...68M}
also gave $\eta$-distributions
for inner, middle, and outer main-belt objects, with peak values
at around 0.95-1.1 for inner, 0.9-1.0 for middle, and 0.85-0.95 for
outer main-belt asteroids. For the interpretation of our single-band data
it is not possible to determine object-specific $\eta$-values from the
measurements. Therefore, we took the $\eta$-relation by Ref.\cite{2018A&A...612A..85A}:
$\eta$ ($\alpha$) = 0.76 ($\pm$ 0.03) + (0.009 $\pm$ 0.001)\,deg$^{-1}$. This relation is based on the analysis of more than 5000 asteroids, all observed
in two broad bands at 9 and 18 $\mu$m. As both bands are close to the MIRI
F1500W band, we consider this solution as the most appropriate for our
analysis. However, instead of the given parameter error, we use a more
conservative $\eta$-error of $\pm$10\% for the NEATM size and albedo calculations.

For very small asteroids (on the decameter scale) the NEATM model is not well tested. In addition, small objects have a faster rotation\cite{2000Icar..148...12P}. In this case, their surface temperature would be better described by the Isothermal Latitude Model (ILM) or Fast-Rotating-Model (FRM)\cite{1989aste.conf..128L} (see also discussions in Ref.\cite{2002aste.book..205H}). We tested the impact of these model assumptions for the size calculation
of typical main-belt asteroids. Such fast-rotating, nearly isothermal asteroids would require significantly larger sizes to produce the measured flux in MIRI F1500W band. For an object at 2.75\,au heliocentric distance (and 2.14\,au observer-centric
distance) and seen under a typical phase angle
of 23$^{\circ}$ our default NEATM solution translates a 1\,$\mu$Jy flux into a 32\,m source diameter (for a geometric albedo of 0.15).
The corresponding FRM solution gives a size around 55\,m. However, the very small objects are only seen at smaller heliocentric distances (and typically larger phase angles) where the differences between both model solutions are less than 50\%, but with the FRM-sizes being always larger.


\subsection{Properties of the eight known asteroids}

The errors in H-mag, $\eta$, and the measured flux are all considered in the NEATM calculations. An absolute flux error of 6\% (10\%) leads approximately to uncertainty of 3\% (5\%) for the size and 5\% (9\%) for the albedo, while the 10\% higher (lower) $\eta$-value increases (decreases) the size. The dominating sources of uncertainty for the size estimates are the assumptions for $\eta$ and the absolute uncertainty for the fluxes, while for the albedo, the large uncertainty in H-mag drives the final errors. Table~\ref{tbl:8ast} summarizes the NEATM input values and our findings for the eight known asteroids among the serendipitous detections. 

Among all known asteroids, (194793) 2001 YP90 was observed by JWST continuously for the longest period of time (3 hours) and the data shows significant flux variations indicating an elongated
body (see Fig.~\ref{lc_YP90}) with a minimum (maximum) flux of $550\,\mu{\rm{Jy}}$ ($1300\,\mu{\rm{Jy}}$). These flux values translate into diameters of $640\pm43$\,m ($922\pm70$\,m) and albedo values of $0.51^{+0.17}_{-0.14}$ ($0.24^{+0.10}_{-0.07}$). Calculated diameters at minimum and maximum flux are first order estimates for the elongation of the asteroid. Diameters and albedos reported in Table~\ref{tbl:8ast} were calculated using an average flux of 1.03\,$\mu$Jy. Obtained JWST light curve is in a good agreement with the data from ground-based telescopes (see section Follow-up observations). 

Albedo values of asteroid 152630 (1997 GP4) derived from our radiometric analysis ($\mathrm{pV=0.29^{+0.12}_{-0.08}}$) are in agreement with the expected value for S-type asteroids ($0.26\pm0.09$\cite{2014Natur.505..629D}; see section Follow-up observations).

Out of the eight known MBAs, only (472944) 2015 GH28 has a published
radiometric diameter. Ref.\cite{Masiero_2011ApJ...741...68M} used 11 W3-band measurements
(from 15/16 Feb 2010, at r$_{helio}$=2.45\,au, $\Delta$=2.23\,au, $\alpha$=23.8$^{\circ}$, W3 band center at 12\,$\mu$m) to derive a size of 2290$\pm$390\,m, no albedo was determined. This is in good agreement with our findings.


\subsection{Properties of the objects with unknown orbits}

For newly-detected objects we used the previously introduced population-driven constraints on their orbits (see method "Orbit estimations") to transform the measured F1500W fluxes into size estimates. As neither H-mag nor albedo are known, we simply determined the minimal and maximal sizes which would be compatible with the orbital and flux constraints.
The procedure is described by the following steps:

\begin{enumerate}
    \item  We use the calculated Observer-Target (O-T) distance, Sun-Target (S-T) distance, and the
   corresponding S-T-O phase angles from the 3-$\sigma$ uncertainty ellipses
   (if not available, we took the 10-$\sigma$ values, or in the worst cases
   we took the full main-belt region with possible S-T distances between
   1.76 and 3.74\,au, and the corresponding full possible range for O-T and
   S-T-O values.)
    
    \item From the wide ranges of compatible orbit solutions, we determine the possible extreme near and far geometries (in S-T, O-T, and S-T-O). Note that these extreme distances and angles are still compatible with the corresponding JWST-centric solar elongation of the TRAPPIST-1 star for the different observing epochs.
    
    \item The smallest possible asteroid size (compatible with the measured flux at 15\,$\mu$m,
   lower flux boundary) is found at the near geometry under the assumptions of
   (i) a low albedo (leading to an IR-bright object), here we assumed a geometric
   albedo of 0.05; (ii) a low beaming parameter (calculated via the Ref.\cite{2017A&A...603A..55A}
   relation for the given S-T-O$_{near}$ phase angle, and lowered by 10\% as done
   for the known asteroids). The largest possible asteroid size (compatible with the
   measured flux at 15\,$\mu$m, upper flux boundary) is found at the far geometry under
   the assumptions of (i) a high albedo (leading to an IR-faint object), here we
   assumed a geometric albedo of 0.30 (see also \cite{Masiero_2011ApJ...741...68M}
   for an overview of albedo distributions as a function of heliocentric distances);
   (ii) a high beaming parameter (calculated via the Ref.\cite{2017A&A...603A..55A}
   relation for the given S-T-O$_{far}$ phase angle, and increased by 10\% as done
   for the known asteroids).
    
    \item The procedure leads to a final size range for each object, compatible with
   flux and orbit constraints.
    
\end{enumerate}

It is important to note that the derived size range for each object is dominated by the
range of possible geometries. The typical 10\% absolute flux error contributes only
about 5\% to the size uncertainty, and different assumptions for the albedos are
almost negligible in the radiometric size determination. All the properties of the 139 new detections are reported in Supplementary Table 1.

For a validation of the method, we handled the 8 known asteroids in exactly the same ways as the 139 unknown ones. The resulting solutions are shown in Figure\,\ref{fig:proof_of_concept} (based on S-T, O-T, and S-T-O values derived from the 3-$\sigma$ ellipse in orbital properties). The derived size ranges are larger, but agree very well with the solutions given in Table~\ref{tbl:8ast} where their true orbits were used. The asteroids are reported in the same order as in Figure\,\ref{fig:proof_of_concept} and in Table~\ref{tbl:8ast}, i.e., asteroid \#1 is 2011 SG255 and asteroid \#8 is (472944) 2015 GH28.


\subsection{Follow-up observations of the eight known asteroids}

We conducted ground-based follow-up observations of a set of known asteroids in our sample to better characterize their phase curve, colors, and rotation period and amplitude. Observations of (194793) 2001~YP90 and 2021~FR9 were acquired with the 1-m Artemis telescope\cite{2022PASP..134j5001B} of the SPECULOOS network\cite{2018SPIE10700E..1ID} and with the 0.6-m TRAPPIST-North\cite{Jehin_2011} telescope. 
2021~FR9 was observed between 2024 February 02-10 at solar phase angles ranging from $1.8^{\circ}$ to $4.3^{\circ}$. The photometry and magnitude calibration to the Johnson V band was performed using the Photometry Pipeline\cite{2017A&C....18...47M}. 
2001~YP90 was observed between 2024 February 01 and March 09 at solar phase angles ranging from $2.9^{\circ}$ to $20.9^{\circ}$ and included longer observation runs to determine its rotation period. We determined a period of $5.7701 \pm 0.0001$\,h and a relatively large amplitude of $0.87 \pm 0.10$\,mag, indicating an elongated body (Fig.~\ref{lc_YP90}).


A series of exposures with SDSS {\it griz} filters  were also obtained for the asteroids 2001 YP90 and 1997 GP4 on February 16, with the 4.3-m Lowell Discovery Telescope (LDT, previously known as Lowell's Discovery Channel Telescope).\cite{Levine2012} These spectro-photometric observations allowed to determine the taxonomic types for these two bright asteroids. For 2001 YP90 the taxonomic fits RMS values suggest that a K-type is the only good fit to the data. Derived values of albedo from our radiometric analysis (see section "Properties of the eight known asteroids") are also compatible with K-type asteroid. For 1997 GP4 asteroid, Sr-type is the best fit, but S- or Sq-types are close in terms of RMS (see Fig.~\ref{DTC}).


\subsection{On the information content and sensitivity of the size-frequency distribution}

Before interpreting the SFD, we assessed its sensitivity to uncertainties and biases in order to determine the size regime over which its information content can reliably be translated into scientific inferences. First, we developed a framework to adequatley propagate the large uncertainties (Fig.\,3.a.) on the asteroid radii onto the SFD. To this end, we followed the Monte Carlo method\cite{Metropolis1949} and generated an ensemble of 10,000 randomly perturbed diameters for each asteroid following a Gaussian distribution using their respective mean and 1-$\sigma$ uncertainty (Fig.\,3.b.). This provides us with 10,000 SFDs over which we can run the subsequent analysis, as well as derive a median SFD and 1-$\sigma$ uncertainty envelope. We note that due to the transformation from linear (sizes) to log (SFD) space, the uncertainty distribution on the SFD is asymmetrical. This means that the SFD derived from the median sizes does not correspond to the actual median SFD. Therefore, not accounting for the size uncertainties when deriving the SFD estimates can lead to biases when the size uncertainties are important. In the present case, not accounting for the size uncertainty leads to an SFD estimate biased towards larger slope (Extended Data Figure\,\ref{sfd_study}.a.).

Similarly we note that it is pivotal to account for the expected distribution of uncertainties as well as the sample size when building the models to be compared with the SFD. Indeed, standard theoretical models are built assuming that a remarkably large number of asteroids are observed with an exquisite precision on their sizes, which is not true in practice. To highlight that aspect, we offer the following simple case: a theoretical population where all objects have the exact same size leading to a vertical SFD. Due to measurement uncertainty, any observation of this population will always return a spread of value due to uncertainties on each individual size measurements, which will result in a sloped SFD. That apparent slope will be dependent on the uncertainty on the size estimates, and the number of objects detected. Figure\,\ref{sfd_study}.b. further develops this point by comparing the true slope of a SFD to its apparent slope as a function of the measurement uncertainty ($\sigma_D$ assumed here to be proportional to the size---which is somewhat consistent with our JWST observations) for a sample size of 150 asteroids. It shows that for a regime where $\sigma_D/D >10\%$ the apparent slope is systematically shallower. In addition, it shows that the $q=-1.45$ and $q=-0.85$ slopes observed match with true slopes of $q\gtrapprox-2.2$ and $q\gtrapprox-0.95$. We use this slope mapping to debias our SFD and match with theoretical models (Figure\,4).

Second, we investigated observational biases. The first bias of observational origin relates to our detection threshold at $\sim$0.5\,$\mu$Jy. This threshold translates into an observational bias emerging in the 20 to 40-m diameter (Fig.\,3.a.), meaning that the current SFD cannot readily be used beyond 40\,m. On the other end of the size regime, large asteroids crossing the FoV are rare due to their lower occurence rates. Therefore, the SFD derived from their detection is affected by small number statistics. To assess the size threshold above which this occurs, we generated synthetic populations of one million objects with exponents ranging from $q = -2.0$ to $q = -0.7$---$N({>}D) = C D^q$---and drew 10,000 random batches of 150 asteroids to assess the size regime over which the derived SFDs present an adequately small level of variance given our number of detections. 

We find that the size cutoff for the sampling bias is dependent on the exponent, with a size cutoff ranging from 70m for $q = -2.0$ to 1200m for $q = -0.7$ (Extended Data Figure\,\ref{sfd_study}.c. and d.). 

This finding also shed lights on the significant tension between the SFD estimate in the 40 to 100\,m range and the number of ``large'' asteroids found ($N({>}300m) = 9$). Indeed, the likelihood of a sample of 150 asteroids with a $q = -1.45$ to also present $N({>}300m) = 9$ is roughly 1:10,000 (i.e., $\gtrapprox4\sigma$). This provides additional support to the fitting of the SFD with a smoother slope beyond $\sim$100\,m, where a transition is seen.

\subsection{Interpretation of the size-frequency distribution}

Knowing which population has been sampled by our JWST observations
is a key for an interpretation. According to Fig.~\ref{all_ai_nature},
orbits of known faint objects located close to the JWST FoV
are not evenly distributed across the main belt.
Instead, they are associated to selected asteroid families,
namely to Polana, Nysa, Massalia, Koronis2, or Karin,
which are both populous and preferentially close to the ecliptic plane,
similarly as the TRAPPIST-1 star.
See Supplementary Table 1 for details.

The observed SFD of individual families are substantially different
(e.g., Fig.~\ref{sfd_families}).
Some of them are steep down to the observational limit,
which occurs at about 1,000\,m,
depending on the respective heliocentric distance and albedo.
Others exhibit a distinct break at around $5,000\,{\rm m}$,
below which the slope becomes shallower,
with the exponent ${\sim}-1.5$,
characteristic of a collisional equilibrium at km sizes.
This is most likely the result of long-term collisional evolution of families.\cite{Broz_2023,Marsset_2023}
In other words, young families (Massalia, Koronis2, Karin)
have a steep SFD,
while old families (Polana, Nysa, \dots)
have a shallow SFD at sub-km and decameter sizes.


\clearpage

\section*{Extended Data Figures \& Tables}
\vspace{2cm}

\clearpage

\begin{figure}
    \centering
    \includegraphics[width=15cm]{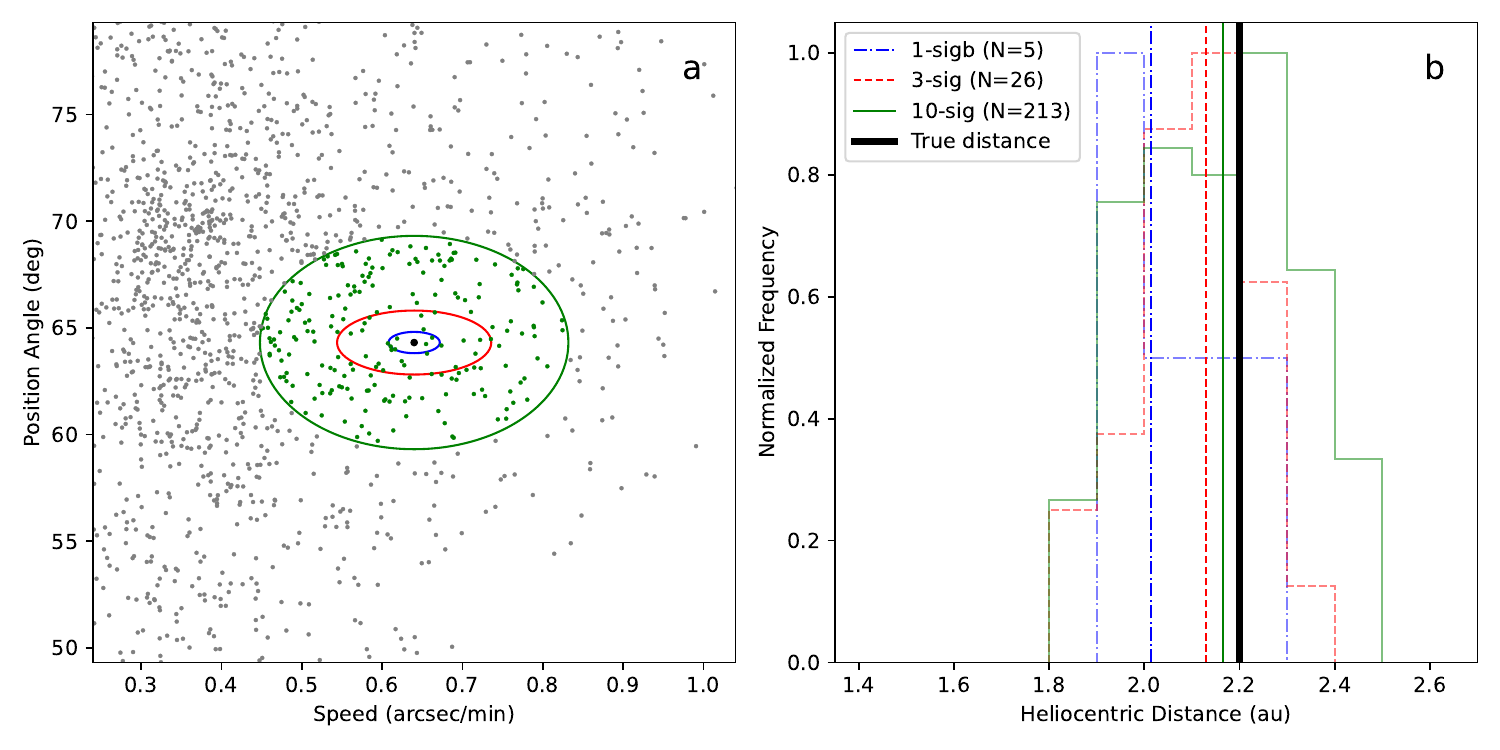}
    \caption{\textbf{Proof-of-concept application of population-based estimation of a distance to an asteroid} \textbf{(a)} Speed and position angle of 2004~GH89 asteroid (black dot) compared to an ensemble of other known asteroids close to the field of view at the time of the observation together with 1-$\sigma$, 3-$\sigma$, and 10-$\sigma$ ellipses (blue, red, green) used to select neighbors. {\bf b.} Histogram of heliocentric distances of known asteroids from 1-$\sigma$, 3-$\sigma$, and 10-$\sigma$ ellipses (blue, red, green). Vertical lines are median heliocentric (S-T) distances for 2004~GH89 asteroid based on the 1-$\sigma$ (N = 5, where N is a number of other known asteroids within the ellipse), 3-$\sigma$ (N = 26), and 10$\sigma$ (N = 213) neighbors showing consistency (i.e., no sensitivity). As the asteroid is known, the derived distance can be compared to the true value (black) as a proof of concept. 
}
    \label{fig:distance_estimation}
\end{figure}

\begin{figure}
\centering
\includegraphics[width=14cm]{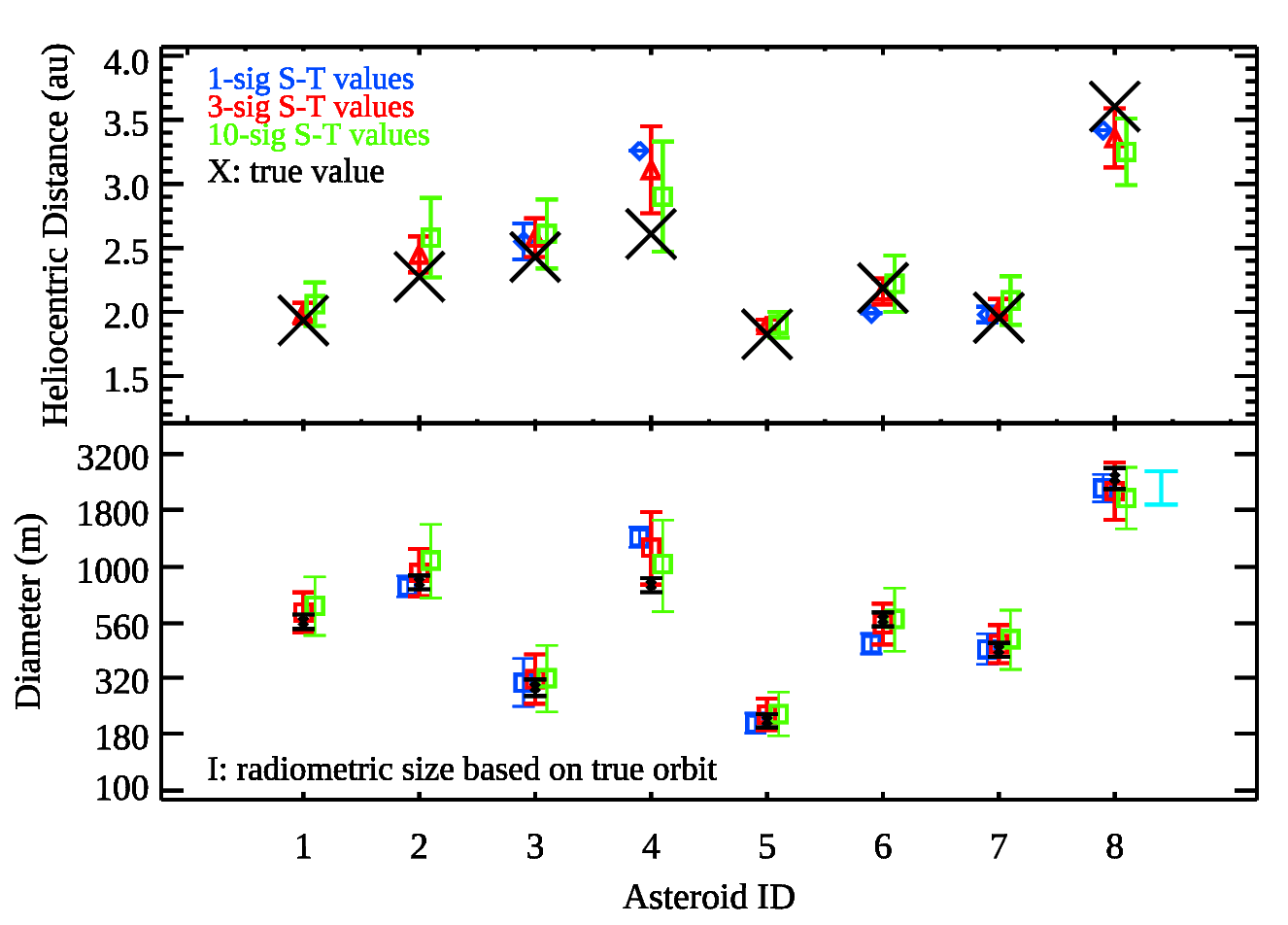}
\caption{
{\bf Proof-of-concept application of the population-based orbit derivation for the eight known asteroids.} (Top) Derived heliocentric distance for the eight known asteroids based on the 1-$\sigma$, 3-$\sigma$, and 10-$\sigma$ neighbors (blue, red, green). (Bottom) Same for the radiometric diameter inferred from the population-based heliocentric distance versus true orbit. Asteroid \#8 has a measured size which is reported in cyan. All 3-$\sigma$- and 10-$\sigma$-neighbors estimates for the distances and the existing size agrees to within 1-$\sigma$ supporting the reliability of the methodology used. See asteroid names in Table~\ref{tbl:8ast} where they are reported in the same order (i.e., asteroid \#1 is 2011 SG255 and asteroid \#8 is (472944) 2015 GH28).
}
\label{fig:proof_of_concept}
\end{figure}

\clearpage

\begin{figure}[h]
\centering
\includegraphics[width=17cm]{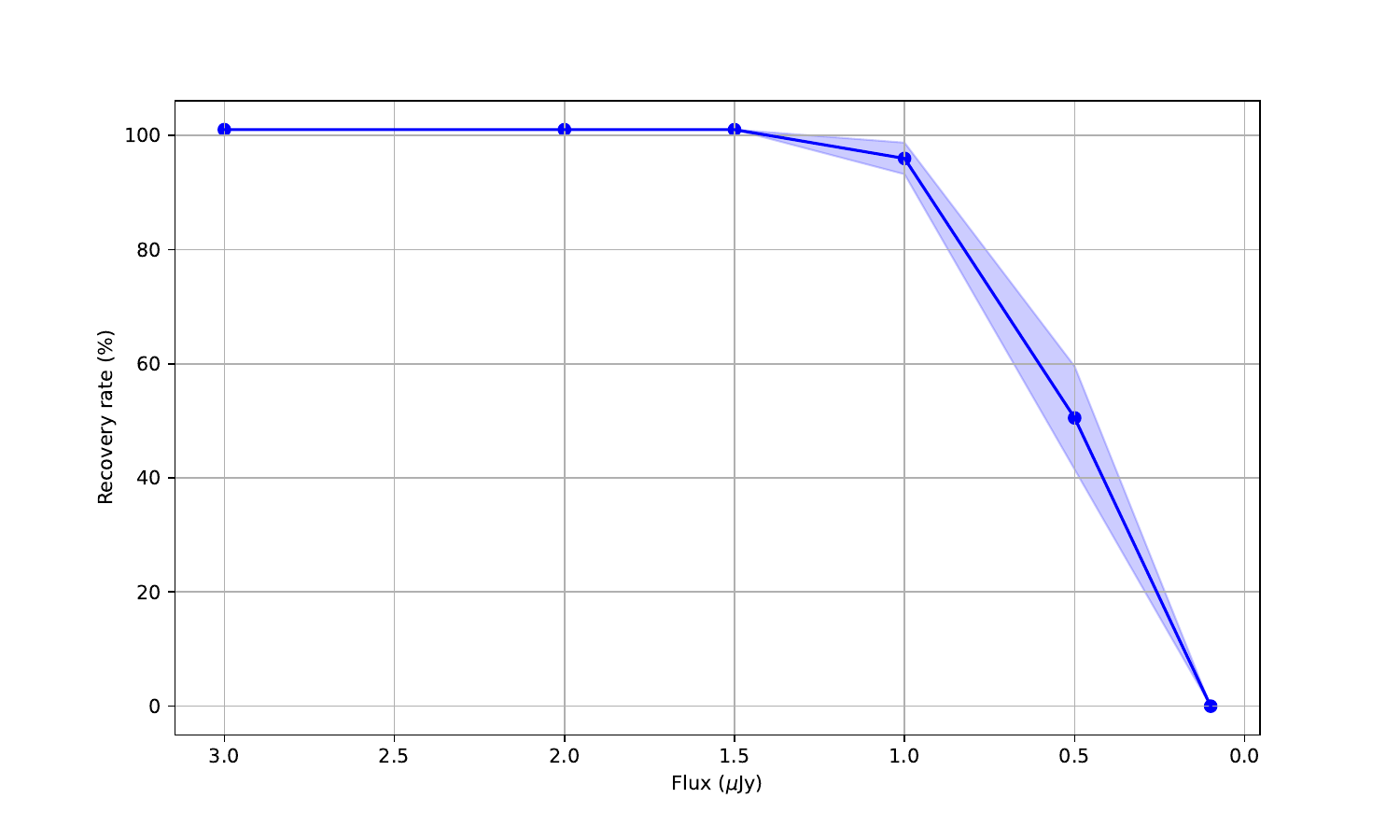}
\caption{ {\bf Completeness test and recovery rate.} Fraction of recovered synthetic asteroids as a function of their flux based on injection recovery tests to assess the completeness of our search and correct the derived size-frequency distribution. The shaded area represent the 1-$\sigma$ deviation from the reported rates seen across a range of injections. The derived cutoff is 0.5$\,\mu$Jy with a observation bias starting at $\sim 1\,\mu$Jy.
}
\label{fig:recovery-rate}
\end{figure}

\clearpage

\begin{figure}[h!]
\centering
\includegraphics[width=15cm]{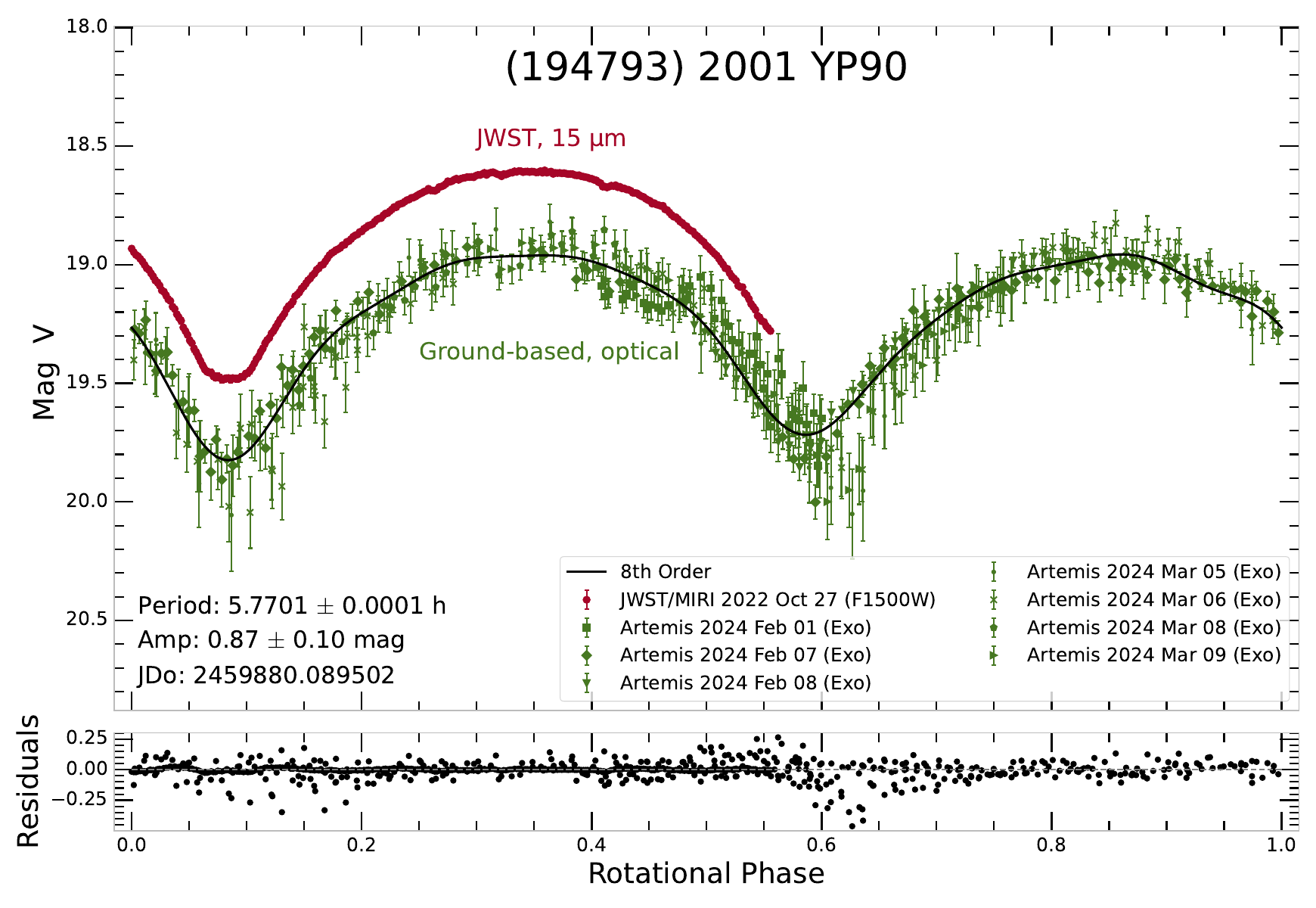}
\caption{
{\bf Phased rotational lightcurve of 2001~YP90.}
Photometric observations of 2001~YP90 obtained with the Artemis telescope\cite{2022PASP..134j5001B} (in green) indicate a rotation period of $5.7701 \pm 0.0001$\,h and an amplitude of $0.87 \pm 0.10$\,mag. The red curve corresponds to the MIRI observations shifted to the V band data (-0.35\,mag for clarity) and shows a very good match with the optical observations in shape and amplitude. The MIRI data uncertainties are plotted but are smaller than the markers size.
}
\label{lc_YP90}
\end{figure}

\clearpage

\begin{figure}[h!]
\centering
\includegraphics[width=17cm]{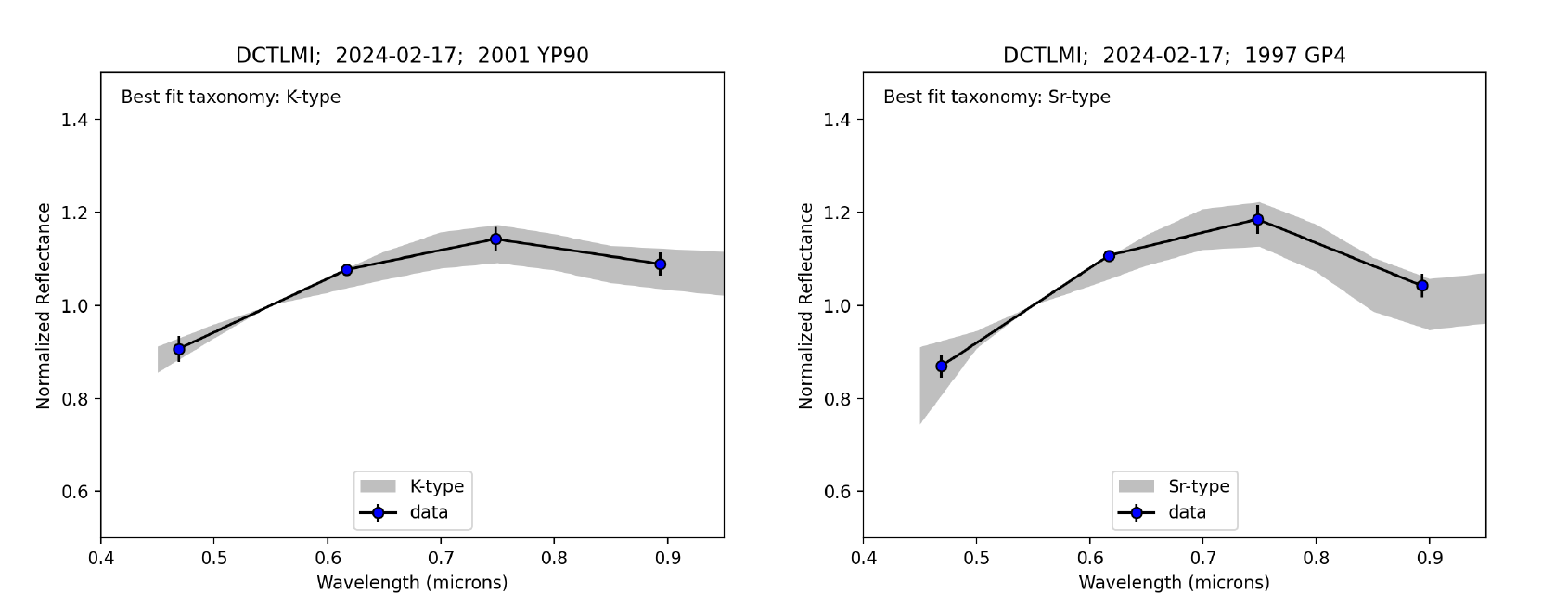}
\caption{
{\bf Spectro-photometric taxonomic types of bright asteroids 2001 YP90 and 1997 GP4.} Data were obtained with SDSS {\it griz} filters. Best-fit taxonomic types were determined based on minimizing RMS residuals between the data and re-sampled templates of taxonomic types in the Bus-DeMeo system. The albedos inferred from our radiometric analysis are in agreement with the ones from taxonomic classifications.}
\label{DTC}
\end{figure}

\clearpage

\begin{figure}
\centering
\vspace*{-40mm}\includegraphics[width=17cm]{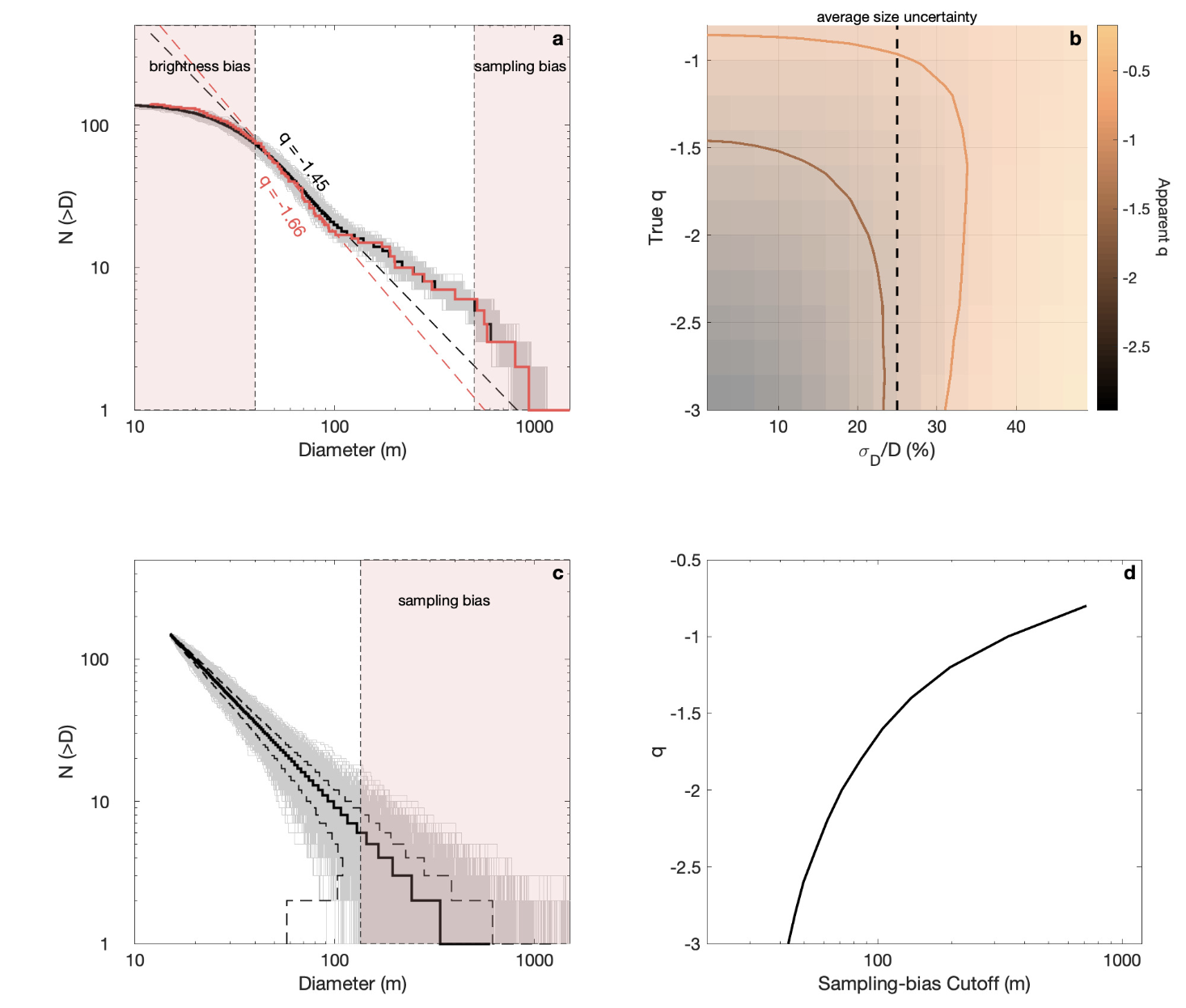}
\vspace*{-25mm}\caption{
{\bf Information content and sensitivity analysis of a size-frequency distribution.}
\textbf{a.} Comparison between the SFD derived using the best size estimate (i.e. median) for each asteroid (red), and the median SFD (black) derived via the Monte Carlo method\cite{Metropolis1949} from an ensemble of 1,000 SFDs generated using an ensemble of 1,000 randomly perturbed diameters for each asteroid. Not accounting for the size uncertainty leads to an SFD estimate biased towards larger slope ($q=-1.66$ vs $q=-1.45$). \textbf{b.}  True versus apparent $q$ exponent for ensembles of 1,000 samples of 150 asteroids drawn from a synthetic family, as a function of the relative measurement uncertainty ($\sigma_{D}/D$). The contours represent the $q=-1.45$ and $q=-0.85$ slopes observed, which match true slopes of $q\gtrapprox-2.2$ and $q\gtrapprox-0.95$ for our average size uncertainty (dashed line). \textbf{c.} SFDs from 3,000 random draws of 150 asteroids from a $q=-1.45$ population of $\sim$1e6 asteroids showing consistent slopes until $\sim$130m due to small-number statistic, the ``sampling-bias cutoff''.  \textbf{c.} Relationship between the sampling-bias cutoff and the q-coefficient of a population for a sample size of 150 asteroids.
}
\label{sfd_study}
\end{figure}

\clearpage

\begin{figure}
\centering
\includegraphics[width=14.5cm]{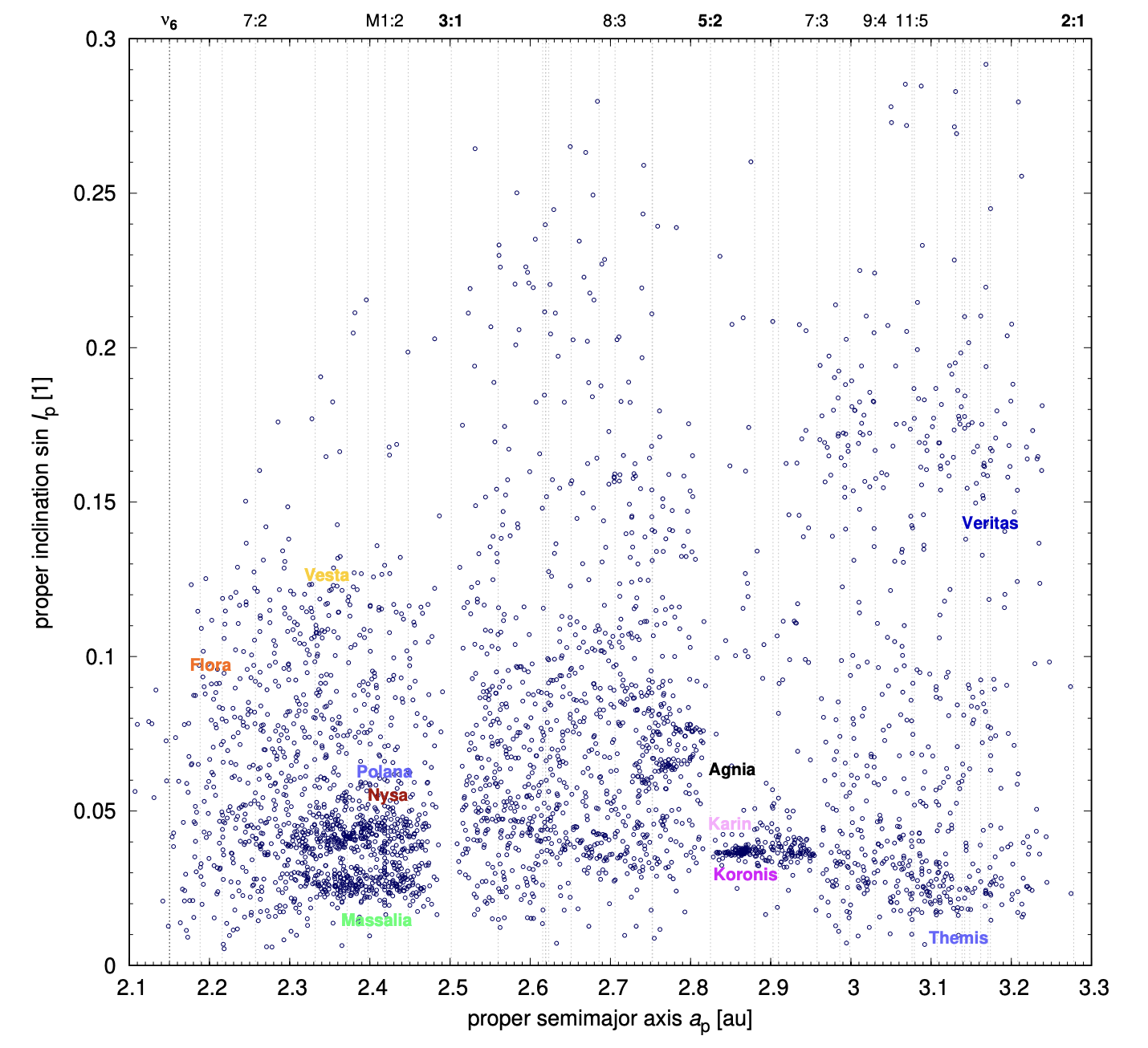}
\caption{
{\bf Orbital elements of known asteroids located close to the field of TRAPPIST-1.}
All these asteroids had a similar proper motion and position angle
as the unknown asteroids observed by the JWST.
Their proper semimajor axis $a_{\rm p}$ versus the proper inclination $\sin i_{\rm p}$
(blue circles)
is compared to other faint main belt asteroids observed by the Catalina Sky Survey\cite{Christensen_2023LPICo2851.2587C}
(gray dots).
Their concentrations ('clouds') correspond to known asteroid families \cite{Nesvorny_2015aste.book..297N}.
Sampling is non-random due to the geometry of JWST observations.
Preferentially, the Nysa, Polana and Massalia families are sampled,
together with other families at low inclinations
(Koronis2, Karin).
}
\label{all_ai_nature}
\end{figure}

\clearpage

\begin{figure}
\centering
\includegraphics[width=12cm]{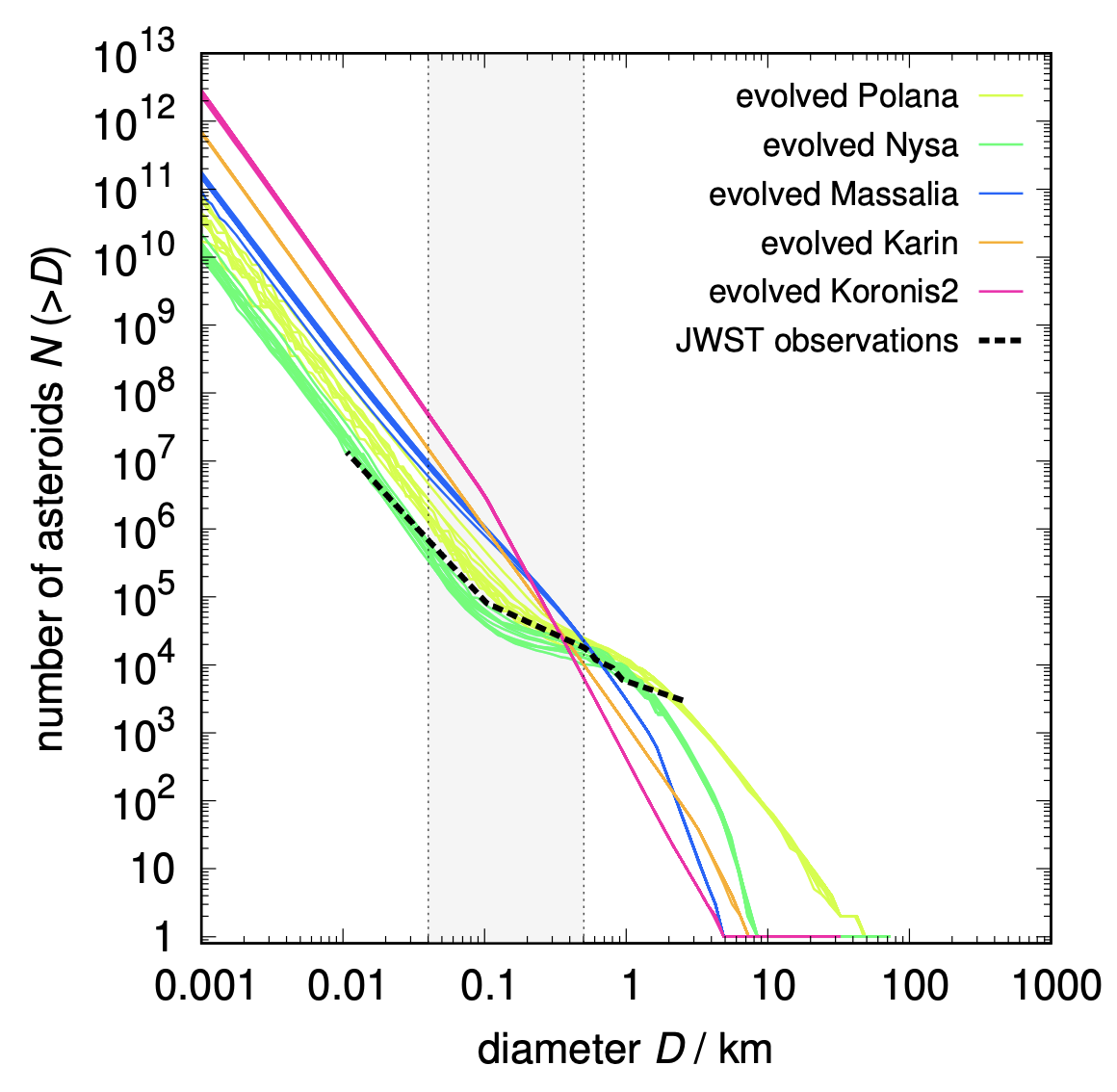}
\caption{
{\bf Young and old asteroid families have very different size-frequency distributions between 1,000 and 50\,m.}
A comparison of synthetic distributions of asteroid families
from refs. \cite{Broz_2023,Marsset_2023,Broz_2024} shows that
prominent young families (Massalia, Koronis2, Karin)
commonly have a steep slope
(the exponent $q \simeq -2.5$ up to $-4$),
while old families (Polana, Nysa)
have a shallow slope ($q \simeq -1$ to $-1.5$).
This difference stems from the fact that hundred-meter-size
bodies are the weakest bodies in terms of their strength
(i.e., energy per unit of mass needed for disruption)
\cite{Benz_1999Icar..142....5B}.
Consequently, their collisional evolution is so substantial
that after approximately 100\,My the exponent changes dramatically
\cite{Broz_2023}.
}
\label{sfd_families}
\end{figure}

\clearpage

\def\tablefoot#1{%
 \par\vspace*{2ex}%
 \parbox{\hsize}{\leftskip0pt\rightskip0pt
 {\noindent\small{\bf Notes.}~#1\par}}%
}

\begin{table}
    \caption{\textbf{Properties of the eight known asteroids.} H$_{fit}$ reports the estimated H magnitudes, r$_{helio}$ the heliocentric distances, $\Delta$ the distance to JWST, $\alpha$ the viewing angle, $\eta$($^*$) the infrared beaming parameter, D the diameters, and pV the albedo.}
    \bigskip
    \hspace*{-20mm}
    \begin{tabular}{c|cccccc|cc}
        Name & H$_{fit}$ & r$_{helio}$(au) & $\Delta$(au) & $\alpha$ & $\eta$($^*$) & flux($\mu$Jy)($^{**}$) & D (m) & pV \\ \hline
2011 SG255 &       19.28 &    1.9328 & 	   1.2798 & 	27.8 & 	 1.01	&   1,660$\pm$100 & 580$\pm$40 &  0.100$\pm^{0.043}_{0.030}$ \\
(152630) 1997 GP4 &  17.09 &    2.2750 &	   1.8284 &	25.2  &	 0.99	&   1,240$\pm$74 & 940$\pm$70 &  0.293 $\pm^{0.115}_{0.084}$\\
2021 FR9   &       19.32 &    2.4289 	&   2.1444 &	24.1  &	 0.98	&   90$\pm$9 & 280$\pm$20 &   0.434$\pm^{0.157}_{0.116}$\\
(194793) 2001 YP90 & 17.32  &   2.6092 	&   1.8337 &	16.7  &	 0.91	&   1,030$\pm$62 & 800$\pm$60 &   0.324$\pm^{0.124}_{0.092}$\\
2013 PG137  &      20.37  &   1.8259 &	   1.3056 &	32.0  &	 1.05	&   200$\pm$12 & 190$\pm$10 &   0.340$\pm^{0.130}_{0.094}$\\
2004 GH89   &      18.15   &  2.1869 	&   1.7275 &	26.3  	& 1.00	&   710$\pm$43 & 560$\pm$40 &   0.308$\pm^{0.119}_{0.088}$\\
2016 UR72  &       18.83  &   1.9551 &	   1.4658 &	29.7  	& 1.03	 &  620$\pm$37 & 400$\pm$30 &   0.322$\pm^{0.126}_{0.090}$\\
(472944) 2015 GH28 & 16.66  &   3.6049 &	   3.2427 &	15.6  	& 0.90	 &  1,630$\pm$163 & 2,500$\pm$270 &   0.061$\pm^{0.030}_{0.019}$\\
    \end{tabular}
    \tablefoot{
    ($^*$) The $\eta$ values were calculated via the $\eta$-relation given above
    \cite{2018A&A...612A..85A}, but we allow for a $\pm$10\% uncertainty.
    The calculated minimum (maximum) $\eta$ values for the 8 asteroids are
    0.81 (1.15).
\newline($^{**}$) The detections of the known asteroids have all very high SNRs, but
    for the size-albedo determination we took an absolute flux error
    of 6\%, covering the MIRI imaging flux calibration, color corrections
    (between the stellar and the asteroid SEDs), and MIRI signal drift
    uncertainties. For the asteroids \#03 (flux below 100$\,\mu$Jy) and \#08 (located in the coronographic part of the MIRI detector) we increased the flux error to 10\%.}
    \label{tbl:8ast}
\end{table}

\clearpage

\section*{Supplementary Information}
\vspace{1cm}
\textbf{Supplementary Table 1} Summary of all the information derived regarding the 139 new asteroids detected (overview shown in Figure\,\ref{SuppT1}).\newline
\textbf{Supplementary Figure 1} Portrait gallery of the 139 new asteroids detected (overview shown in Figure\,\ref{SuppF1}).

\clearpage
\begin{figure}
\centering
\vspace*{-30mm}\hspace*{-00mm}\includegraphics[width=15cm]{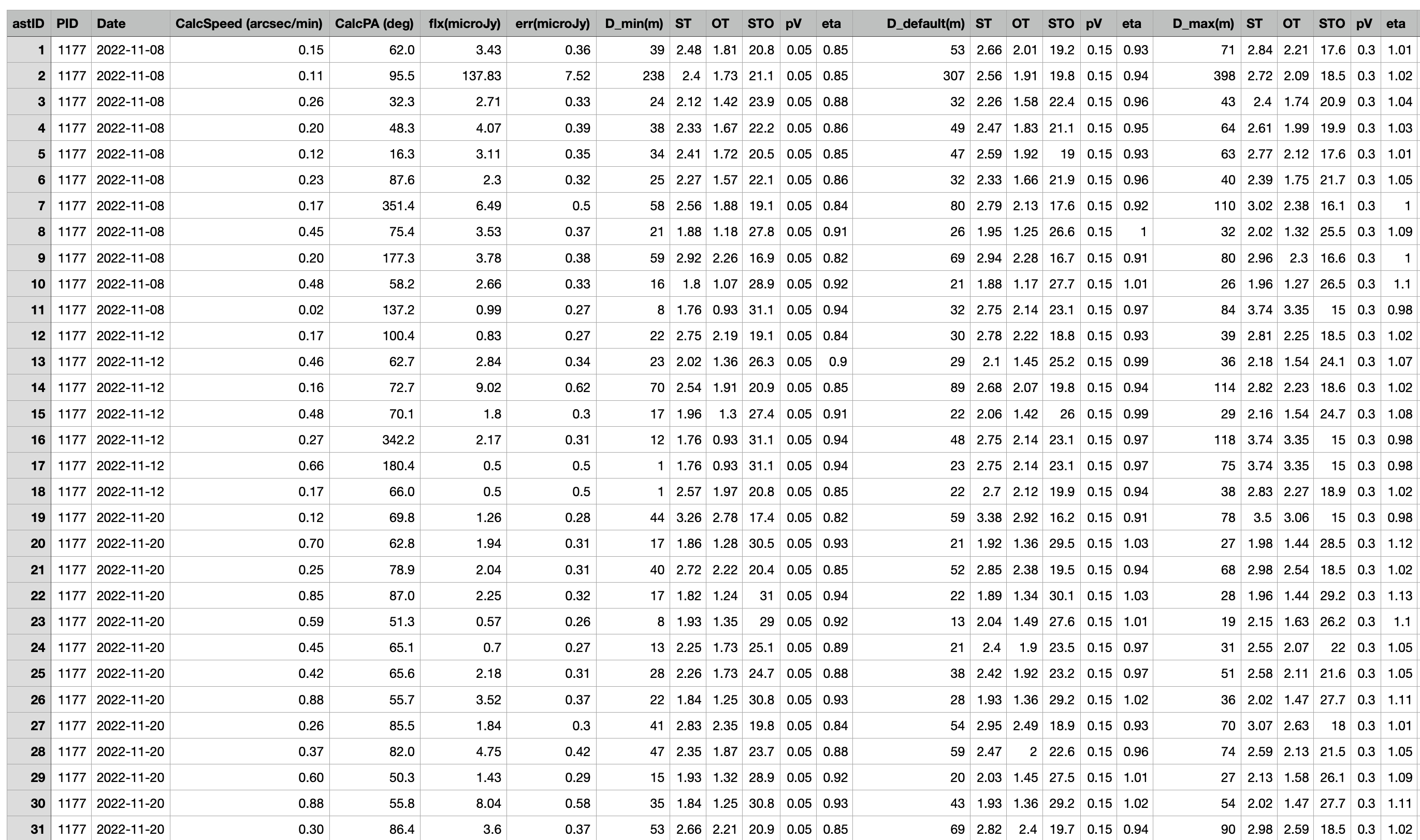}
\caption{
{\bf Overiew of the Supplementary Table 1}
Supplementary Table 1 presents all the information gathered regarding the 139 detections of unknown asteroids by JWST presented here.
}
\label{SuppT1}
\end{figure}

\clearpage
\begin{figure}
\centering
\vspace*{-30mm}\hspace*{-00mm}\includegraphics[width=15cm]{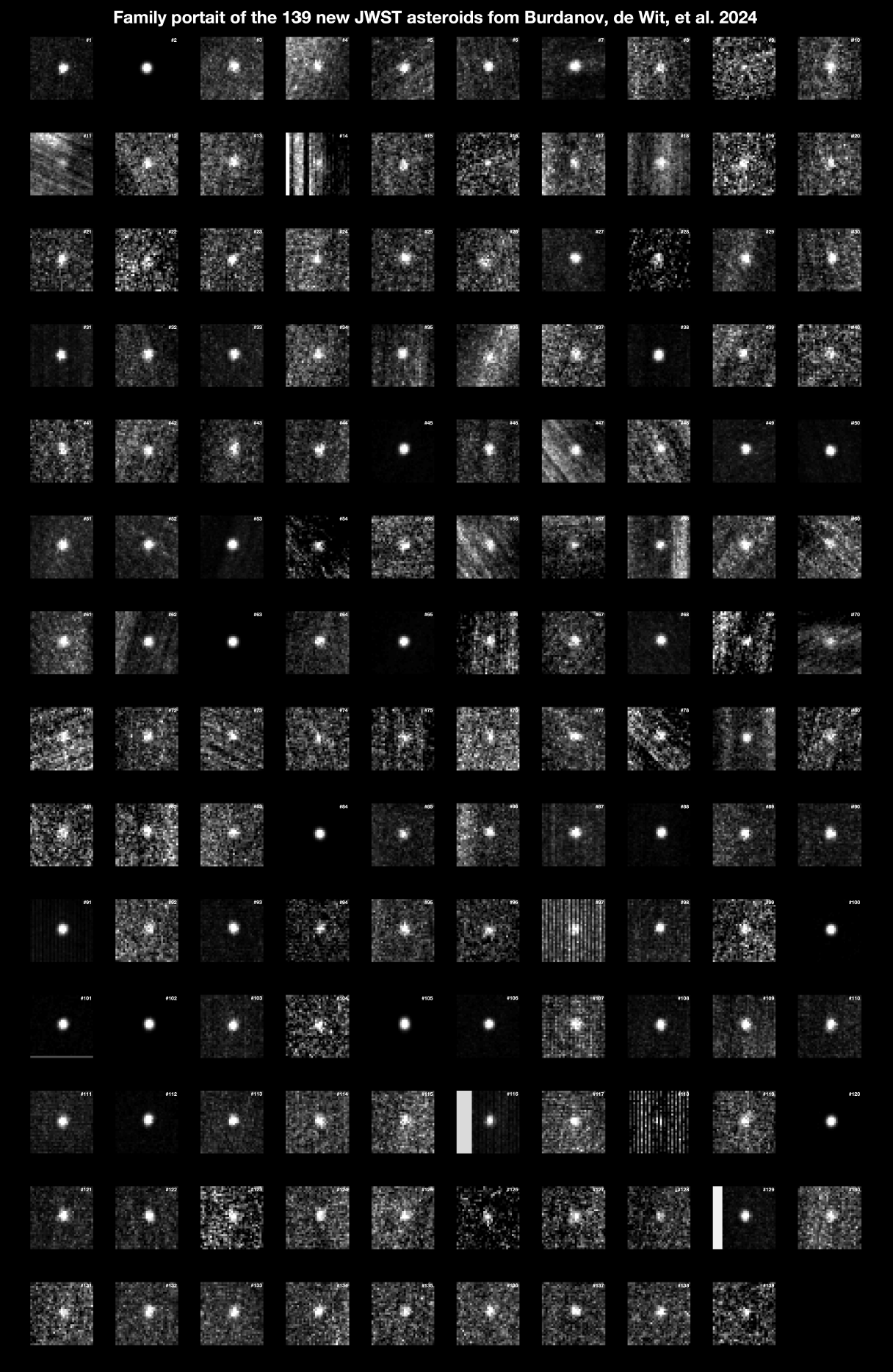}
\caption{
{\bf Overiew of the Supplementary Figure 1}
Supplementary Figure 1 presents a series of snapshot of the 139 detections of unknown asteroids by JWST presented here.
}
\label{SuppF1}
\end{figure}

\clearpage

\end{methods}

\newpage
\pagebreak
\clearpage

\section*{References}
\bibliography{References.bib}

\begin{thebibliography}{10}
\expandafter\ifx\csname url\endcsname\relax
  \def\url#1{\texttt{#1}}\fi
\expandafter\ifx\csname urlprefix\endcsname\relax\def\urlprefix{URL }\fi
\providecommand{\bibinfo}[2]{#2}
\providecommand{\eprint}[2][]{\url{#2}}

\bibitem{Cheng_2018P&SS..157..104C}
\bibinfo{author}{{Cheng}, A.~F.} \emph{et~al.}
\newblock \bibinfo{title}{{AIDA DART asteroid deflection test: Planetary
  defense and science objectives}}.
\newblock \emph{\bibinfo{journal}{\planss}} \textbf{\bibinfo{volume}{157}},
  \bibinfo{pages}{104--115} (\bibinfo{year}{2018}).

\bibitem{Brown2002}
\bibinfo{author}{{Brown}, P.}, \bibinfo{author}{{Spalding}, R.~E.},
  \bibinfo{author}{{ReVelle}, D.~O.}, \bibinfo{author}{{Tagliaferri}, E.} \&
  \bibinfo{author}{{Worden}, S.~P.}
\newblock \bibinfo{title}{{The flux of small near-Earth objects colliding with
  the Earth}}.
\newblock \emph{\bibinfo{journal}{\nat}} \textbf{\bibinfo{volume}{420}},
  \bibinfo{pages}{294--296} (\bibinfo{year}{2002}).

\bibitem{Chyba_1993Natur.361...40C}
\bibinfo{author}{{Chyba}, C.~F.}, \bibinfo{author}{{Thomas}, P.~J.} \&
  \bibinfo{author}{{Zahnle}, K.~J.}
\newblock \bibinfo{title}{{The 1908 Tunguska explosion: atmospheric disruption
  of a stony asteroid}}.
\newblock \emph{\bibinfo{journal}{\nat}} \textbf{\bibinfo{volume}{361}},
  \bibinfo{pages}{40--44} (\bibinfo{year}{1993}).

\bibitem{Chao_1960Sci...132..220C}
\bibinfo{author}{{Chao}, E.~C.~T.}, \bibinfo{author}{{Shoemaker}, E.~M.} \&
  \bibinfo{author}{{Madsen}, B.~M.}
\newblock \bibinfo{title}{{First Natural Occurrence of Coesite}}.
\newblock \emph{\bibinfo{journal}{Science}} \textbf{\bibinfo{volume}{132}},
  \bibinfo{pages}{220--222} (\bibinfo{year}{1960}).

\bibitem{Stoffler_2002M&PS...37.1893S}
\bibinfo{author}{{St{\"o}ffler}, D.}, \bibinfo{author}{{Artemieva}, N.~A.} \&
  \bibinfo{author}{{Pierazzo}, E.}
\newblock \bibinfo{title}{{Modeling the Ries-Steinheim impact event and the
  formation of the moldavite strewn field}}.
\newblock \emph{\bibinfo{journal}{\maps}} \textbf{\bibinfo{volume}{37}},
  \bibinfo{pages}{1893--1907} (\bibinfo{year}{2002}).

\bibitem{Reddy2019}
\bibinfo{author}{{Reddy}, V.} \emph{et~al.}
\newblock \bibinfo{title}{{Near-Earth asteroid 2012 TC4 observing campaign:
  Results from a global planetary defense exercise}}.
\newblock \emph{\bibinfo{journal}{\icarus}} \textbf{\bibinfo{volume}{326}},
  \bibinfo{pages}{133--150} (\bibinfo{year}{2019}).

\bibitem{Bottke_2005Icar..175..111B}
\bibinfo{author}{{Bottke}, W.~F.} \emph{et~al.}
\newblock \bibinfo{title}{{The fossilized size distribution of the main
  asteroid belt}}.
\newblock \emph{\bibinfo{journal}{\icarus}} \textbf{\bibinfo{volume}{175}},
  \bibinfo{pages}{111--140} (\bibinfo{year}{2005}).

\bibitem{Farinella_1998Icar..132..378F}
\bibinfo{author}{{Farinella}, P.}, \bibinfo{author}{{Vokrouhlick{\'y}}, D.} \&
  \bibinfo{author}{{Hartmann}, W.~K.}
\newblock \bibinfo{title}{{Meteorite Delivery via Yarkovsky Orbital Drift}}.
\newblock \emph{\bibinfo{journal}{\icarus}} \textbf{\bibinfo{volume}{132}},
  \bibinfo{pages}{378--387} (\bibinfo{year}{1998}).

\bibitem{Chesley_2003Sci...302.1739C}
\bibinfo{author}{{Chesley}, S.~R.} \emph{et~al.}
\newblock \bibinfo{title}{{Direct Detection of the Yarkovsky Effect by Radar
  Ranging to Asteroid 6489 Golevka}}.
\newblock \emph{\bibinfo{journal}{Science}} \textbf{\bibinfo{volume}{302}},
  \bibinfo{pages}{1739--1742} (\bibinfo{year}{2003}).

\bibitem{Brown_2013Natur.503..238B}
\bibinfo{author}{{Brown}, P.~G.} \emph{et~al.}
\newblock \bibinfo{title}{{A 500-kiloton airburst over Chelyabinsk and an
  enhanced hazard from small impactors}}.
\newblock \emph{\bibinfo{journal}{\nat}} \textbf{\bibinfo{volume}{503}},
  \bibinfo{pages}{238--241} (\bibinfo{year}{2013}).

\bibitem{Harris_2021Icar..36514452H}
\bibinfo{author}{{Harris}, A.~W.} \& \bibinfo{author}{{Chodas}, P.~W.}
\newblock \bibinfo{title}{{The population of near-earth asteroids revisited and
  updated}}.
\newblock \emph{\bibinfo{journal}{\icarus}} \textbf{\bibinfo{volume}{365}},
  \bibinfo{pages}{114452} (\bibinfo{year}{2021}).

\bibitem{Nesvorny_2024Icar..41115922N}
\bibinfo{author}{{Nesvorn{\'y}}, D.} \emph{et~al.}
\newblock \bibinfo{title}{{NEOMOD 2: An updated model of Near-Earth Objects
  from a decade of Catalina Sky Survey observations}}.
\newblock \emph{\bibinfo{journal}{\icarus}} \textbf{\bibinfo{volume}{411}},
  \bibinfo{pages}{115922} (\bibinfo{year}{2024}).
\newblock \eprint{2312.09406}.

\bibitem{Broz_2023}
\bibinfo{author}{{Bro\v{z}}, M.} \emph{et~al.}
\newblock \bibinfo{title}{{Young asteroid families as the primary source of
  meteorites}}.
\newblock \emph{\bibinfo{journal}{Nature, submit.}}  (\bibinfo{year}{2024}).
\newblock \eprint{2403.08552}.

\bibitem{Marsset_2023}
\bibinfo{author}{{Marsset}, M.} \emph{et~al.}
\newblock \bibinfo{title}{{The Massalia asteroid family as the origin of
  ordinary L chondrites}}.
\newblock \emph{\bibinfo{journal}{Nature, submit.}}  (\bibinfo{year}{2024}).
\newblock \eprint{2403.08548}.

\bibitem{Mueller2023A&A}
\bibinfo{author}{{M{\"u}ller}, T.~G.} \emph{et~al.}
\newblock \bibinfo{title}{{Asteroids seen by JWST-MIRI: Radiometric size,
  distance, and orbit constraints}}.
\newblock \emph{\bibinfo{journal}{\aap}} \textbf{\bibinfo{volume}{670}},
  \bibinfo{pages}{A53} (\bibinfo{year}{2023}).
\newblock \eprint{2302.06921}.

\bibitem{tyson1992limits}
\bibinfo{author}{Tyson, J.}, \bibinfo{author}{Guhathakurta, P.},
  \bibinfo{author}{Bernstein, G.} \& \bibinfo{author}{Hut, P.}
\newblock \bibinfo{title}{Limits on the surface density of faint kuiper belt
  objects}.
\newblock In \emph{\bibinfo{booktitle}{American Astronomical Society Meeting
  Abstracts}}, vol. \bibinfo{volume}{181}, \bibinfo{pages}{06--10}
  (\bibinfo{year}{1992}).

\bibitem{shao2014finding}
\bibinfo{author}{Shao, M.} \emph{et~al.}
\newblock \bibinfo{title}{Finding very small near-earth asteroids using
  synthetic tracking}.
\newblock \emph{\bibinfo{journal}{\apj}} \textbf{\bibinfo{volume}{782}},
  \bibinfo{pages}{1} (\bibinfo{year}{2014}).

\bibitem{Burdanov2023MNRAS}
\bibinfo{author}{{Burdanov}, A.~Y.}, \bibinfo{author}{{Hasler}, S.~N.} \&
  \bibinfo{author}{{de Wit}, J.}
\newblock \bibinfo{title}{{GPU-based framework for detecting small Solar system
  bodies in targeted exoplanet surveys}}.
\newblock \emph{\bibinfo{journal}{\mnras}} \textbf{\bibinfo{volume}{521}},
  \bibinfo{pages}{4568--4578} (\bibinfo{year}{2023}).
\newblock \eprint{2303.07293}.

\bibitem{Nesvorny_2015aste.book..297N}
\bibinfo{author}{{Nesvorn{\'y}}, D.}, \bibinfo{author}{{Bro{\v z}}, M.} \&
  \bibinfo{author}{{Carruba}, V.}
\newblock \bibinfo{title}{{Identification and Dynamical Properties of Asteroid
  Families}}.
\newblock In \bibinfo{editor}{{Michel}, P.}, \bibinfo{editor}{{DeMeo}, F.~E.}
  \& \bibinfo{editor}{{Bottke}, W.~F.} (eds.)
  \emph{\bibinfo{booktitle}{Asteroids IV}}, \bibinfo{pages}{297--321}
  (\bibinfo{publisher}{Univ. Arizona Press}, \bibinfo{year}{2015}).

\bibitem{Gladman1997A&A}
\bibinfo{author}{{Gladman}, B.} \& \bibinfo{author}{{Kavelaars}, J.~J.}
\newblock \bibinfo{title}{{Kuiper Belt searches from the Palomar 5-m
  telescope.}}
\newblock \emph{\bibinfo{journal}{\aap}} \textbf{\bibinfo{volume}{317}},
  \bibinfo{pages}{L35--L38} (\bibinfo{year}{1997}).
\newblock \eprint{astro-ph/9610150}.

\bibitem{Bernstein2004AJ}
\bibinfo{author}{{Bernstein}, G.~M.} \emph{et~al.}
\newblock \bibinfo{title}{{The Size Distribution of Trans-Neptunian Bodies}}.
\newblock \emph{\bibinfo{journal}{\aj}} \textbf{\bibinfo{volume}{128}},
  \bibinfo{pages}{1364--1390} (\bibinfo{year}{2004}).
\newblock \eprint{astro-ph/0308467}.

\bibitem{Zhai2014ApJ}
\bibinfo{author}{{Zhai}, C.} \emph{et~al.}
\newblock \bibinfo{title}{{Detection of a Faint Fast-moving Near-Earth Asteroid
  Using the Synthetic Tracking Technique}}.
\newblock \emph{\bibinfo{journal}{\apj}} \textbf{\bibinfo{volume}{792}},
  \bibinfo{pages}{60} (\bibinfo{year}{2014}).
\newblock \eprint{1403.4353}.

\bibitem{Heinze2015AJ}
\bibinfo{author}{{Heinze}, A.~N.}, \bibinfo{author}{{Metchev}, S.} \&
  \bibinfo{author}{{Trollo}, J.}
\newblock \bibinfo{title}{{Digital Tracking Observations Can Discover Asteroids
  10 Times Fainter Than Conventional Searches}}.
\newblock \emph{\bibinfo{journal}{\aj}} \textbf{\bibinfo{volume}{150}},
  \bibinfo{pages}{125} (\bibinfo{year}{2015}).
\newblock \eprint{1508.01599}.

\bibitem{2023MNRAS.526.3601H}
\bibinfo{author}{{Hasler}, S.~N.} \emph{et~al.}
\newblock \bibinfo{title}{{Small body harvest with the Antarctic Search for
  Transiting Exoplanets (ASTEP) project}}.
\newblock \emph{\bibinfo{journal}{\mnras}} \textbf{\bibinfo{volume}{526}},
  \bibinfo{pages}{3601--3609} (\bibinfo{year}{2023}).
\newblock \eprint{2309.14180}.

\bibitem{Rieke2015}
\bibinfo{author}{{Rieke}, G.~H.} \emph{et~al.}
\newblock \bibinfo{title}{{The Mid-Infrared Instrument for the James Webb Space
  Telescope, VII: The MIRI Detectors}}.
\newblock \emph{\bibinfo{journal}{\pasp}} \textbf{\bibinfo{volume}{127}},
  \bibinfo{pages}{665} (\bibinfo{year}{2015}).
\newblock \eprint{1508.02362}.

\bibitem{Dohnanyi_1969JGR....74.2531D}
\bibinfo{author}{{Dohnanyi}, J.~S.}
\newblock \bibinfo{title}{{Collisional Model of Asteroids and Their Debris}}.
\newblock \emph{\bibinfo{journal}{\jgr}} \textbf{\bibinfo{volume}{74}},
  \bibinfo{pages}{2531--2554} (\bibinfo{year}{1969}).

\bibitem{Benz_1999Icar..142....5B}
\bibinfo{author}{{Benz}, W.} \& \bibinfo{author}{{Asphaug}, E.}
\newblock \bibinfo{title}{{Catastrophic Disruptions Revisited}}.
\newblock \emph{\bibinfo{journal}{\icarus}} \textbf{\bibinfo{volume}{142}},
  \bibinfo{pages}{5--20} (\bibinfo{year}{1999}).
\newblock \eprint{arXiv:astro-ph/9907117}.

\bibitem{Broz_2024}
\bibinfo{author}{{Bro\v{z}}, M.} \emph{et~al.}
\newblock \bibinfo{title}{{Source regions of carbonaceous meteorites and
  NEOs}}.
\newblock \emph{\bibinfo{journal}{\aap, submit.}}  (\bibinfo{year}{2024}).

\bibitem{Gladman_2009Icar..202..104G}
\bibinfo{author}{{Gladman}, B.~J.} \emph{et~al.}
\newblock \bibinfo{title}{{On the asteroid belt's orbital and size
  distribution}}.
\newblock \emph{\bibinfo{journal}{\icarus}} \textbf{\bibinfo{volume}{202}},
  \bibinfo{pages}{104--118} (\bibinfo{year}{2009}).

\bibitem{Ryan_2015A&A...578A..42R}
\bibinfo{author}{{Ryan}, E.~L.} \emph{et~al.}
\newblock \bibinfo{title}{{The kilometer-sized Main Belt asteroid population
  revealed by Spitzer}}.
\newblock \emph{\bibinfo{journal}{\aap}} \textbf{\bibinfo{volume}{578}},
  \bibinfo{pages}{A42} (\bibinfo{year}{2015}).

\bibitem{Garcia_2024A&A...683A.122G}
\bibinfo{author}{{Garc{\'\i}a-Mart{\'\i}n}, P.} \emph{et~al.}
\newblock \bibinfo{title}{{Hubble Asteroid Hunter. III. Physical properties of
  newly found asteroids}}.
\newblock \emph{\bibinfo{journal}{\aap}} \textbf{\bibinfo{volume}{683}},
  \bibinfo{pages}{A122} (\bibinfo{year}{2024}).
\newblock \eprint{2401.02605}.

\bibitem{Redfield2024}
\bibinfo{author}{{Redfield}, S.} \emph{et~al.}
\newblock \bibinfo{title}{{Report of the Working Group on Strategic Exoplanet
  Initiatives with HST and JWST}}.
\newblock \emph{\bibinfo{journal}{arXiv e-prints}}
  \bibinfo{pages}{arXiv:2404.02932} (\bibinfo{year}{2024}).
\newblock \eprint{2404.02932}.

\bibitem{Greene_2023Natur.618...39G}
\bibinfo{author}{{Greene}, T.~P.} \emph{et~al.}
\newblock \bibinfo{title}{{Thermal emission from the Earth-sized exoplanet
  TRAPPIST-1 b using JWST}}.
\newblock \emph{\bibinfo{journal}{\nat}} \textbf{\bibinfo{volume}{618}},
  \bibinfo{pages}{39--42} (\bibinfo{year}{2023}).
\newblock \eprint{2303.14849}.

\bibitem{Zieba_2023Natur.620..746Z}
\bibinfo{author}{{Zieba}, S.} \emph{et~al.}
\newblock \bibinfo{title}{{No thick carbon dioxide atmosphere on the rocky
  exoplanet TRAPPIST-1 c}}.
\newblock \emph{\bibinfo{journal}{\nat}} \textbf{\bibinfo{volume}{620}},
  \bibinfo{pages}{746--749} (\bibinfo{year}{2023}).
\newblock \eprint{2306.10150}.

\bibitem{Farinella1992}
\bibinfo{author}{{Farinella}, P.}, \bibinfo{author}{{Davis}, D.~R.},
  \bibinfo{author}{{Paolicchi}, P.}, \bibinfo{author}{{Cellino}, A.} \&
  \bibinfo{author}{{Zappala}, V.}
\newblock \bibinfo{title}{{Asteroid collisional evolution - an integrated model
  for the evolution of asteroid rotation rates}}.
\newblock \emph{\bibinfo{journal}{\aap}} \textbf{\bibinfo{volume}{253}},
  \bibinfo{pages}{604--614} (\bibinfo{year}{1992}).

\bibitem{Bottke2006}
\bibinfo{author}{{Bottke}, J., William~F.},
  \bibinfo{author}{{Vokrouhlick{\'y}}, D.}, \bibinfo{author}{{Rubincam}, D.~P.}
  \& \bibinfo{author}{{Nesvorn{\'y}}, D.}
\newblock \bibinfo{title}{{The Yarkovsky and Yorp Effects: Implications for
  Asteroid Dynamics}}.
\newblock \emph{\bibinfo{journal}{Annual Review of Earth and Planetary
  Sciences}} \textbf{\bibinfo{volume}{34}}, \bibinfo{pages}{157--191}
  (\bibinfo{year}{2006}).

\bibitem{Carruba2020}
\bibinfo{author}{{Carruba}, V.} \emph{et~al.}
\newblock \bibinfo{title}{{The population of rotational fission clusters inside
  asteroid collisional families}}.
\newblock \emph{\bibinfo{journal}{Nature Astronomy}}
  \textbf{\bibinfo{volume}{4}}, \bibinfo{pages}{83--88} (\bibinfo{year}{2020}).

\bibitem{Polishook2014}
\bibinfo{author}{{Polishook}, D.}
\newblock \bibinfo{title}{{Spin axes and shape models of asteroid pairs:
  Fingerprints of YORP and a path to the density of rubble piles}}.
\newblock \emph{\bibinfo{journal}{\icarus}} \textbf{\bibinfo{volume}{241}},
  \bibinfo{pages}{79--96} (\bibinfo{year}{2014}).
\newblock \eprint{1406.3359}.

\bibitem{Dinsmore2023}
\bibinfo{author}{{Dinsmore}, J.~T.} \& \bibinfo{author}{{de Wit}, J.}
\newblock \bibinfo{title}{{Constraining the interiors of asteroids through
  close encounters}}.
\newblock \emph{\bibinfo{journal}{\mnras}} \textbf{\bibinfo{volume}{520}},
  \bibinfo{pages}{3459--3475} (\bibinfo{year}{2023}).
\newblock \eprint{2210.10754}.

\bibitem{LSST_2009arXiv0912.0201L}
\bibinfo{author}{{LSST Science Collaboration}} \emph{et~al.}
\newblock \bibinfo{title}{{LSST Science Book, Version 2.0}}.
\newblock \emph{\bibinfo{journal}{arXiv e-prints}}
  \bibinfo{pages}{arXiv:0912.0201} (\bibinfo{year}{2009}).
\newblock \eprint{0912.0201}.

\bibitem{Alvarez1980}
\bibinfo{author}{{Alvarez}, L.~W.}, \bibinfo{author}{{Alvarez}, W.},
  \bibinfo{author}{{Asaro}, F.} \& \bibinfo{author}{{Michel}, H.~V.}
\newblock \bibinfo{title}{{Extraterrestrial Cause for the Cretaceous-Tertiary
  Extinction}}.
\newblock \emph{\bibinfo{journal}{Science}} \textbf{\bibinfo{volume}{208}},
  \bibinfo{pages}{1095--1108} (\bibinfo{year}{1980}).

\bibitem{Metropolis1949}
\bibinfo{author}{Metropolis, N.} \& \bibinfo{author}{Ulam, S.}
\newblock \bibinfo{title}{The monte carlo method}.
\newblock \emph{\bibinfo{journal}{Journal of the American Statistical
  Association}} \textbf{\bibinfo{volume}{44}}, \bibinfo{pages}{335--341}
  (\bibinfo{year}{1949}).
\newblock \urlprefix\url{http://www.jstor.org/stable/2280232}.

\bibitem{harris2020array}
\bibinfo{author}{Harris, C.~R.} \emph{et~al.}
\newblock \bibinfo{title}{Array programming with {NumPy}}.
\newblock \emph{\bibinfo{journal}{\nat}} \textbf{\bibinfo{volume}{585}},
  \bibinfo{pages}{357--362} (\bibinfo{year}{2020}).
\newblock \urlprefix\url{https://doi.org/10.1038/s41586-020-2649-2}.

\bibitem{Hunter:2007}
\bibinfo{author}{Hunter, J.~D.}
\newblock \bibinfo{title}{Matplotlib: A 2d graphics environment}.
\newblock \emph{\bibinfo{journal}{Computing in Science \& Engineering}}
  \textbf{\bibinfo{volume}{9}}, \bibinfo{pages}{90--95} (\bibinfo{year}{2007}).

\bibitem{2013A&A...558A..33A}
\bibinfo{author}{{Astropy Collaboration}} \emph{et~al.}
\newblock \bibinfo{title}{{Astropy: A community Python package for astronomy}}.
\newblock \emph{\bibinfo{journal}{\aap}} \textbf{\bibinfo{volume}{558}},
  \bibinfo{pages}{A33} (\bibinfo{year}{2013}).
\newblock \eprint{1307.6212}.

\bibitem{2018AJ....156..123A}
\bibinfo{author}{{Astropy Collaboration}} \emph{et~al.}
\newblock \bibinfo{title}{{The Astropy Project: Building an Open-science
  Project and Status of the v2.0 Core Package}}.
\newblock \emph{\bibinfo{journal}{\aj}} \textbf{\bibinfo{volume}{156}},
  \bibinfo{pages}{123} (\bibinfo{year}{2018}).
\newblock \eprint{1801.02634}.

\bibitem{2020SciPy-NMeth}
\bibinfo{author}{Virtanen, P.} \emph{et~al.}
\newblock \bibinfo{title}{{{SciPy} 1.0: Fundamental Algorithms for Scientific
  Computing in Python}}.
\newblock \emph{\bibinfo{journal}{Nature Methods}}
  \textbf{\bibinfo{volume}{17}}, \bibinfo{pages}{261--272}
  (\bibinfo{year}{2020}).

\bibitem{reback2020pandas}
\bibinfo{author}{development team, T.~P.}
\newblock \bibinfo{title}{pandas-dev/pandas: Pandas} (\bibinfo{year}{2020}).
\newblock \urlprefix\url{https://doi.org/10.5281/zenodo.3509134}.

\bibitem{mckinney-proc-scipy-2010}
\bibinfo{author}{{W}es {M}c{K}inney}.
\newblock \bibinfo{title}{{D}ata {S}tructures for {S}tatistical {C}omputing in
  {P}ython}.
\newblock In \bibinfo{editor}{{S}t\'efan van~der {W}alt} \&
  \bibinfo{editor}{{J}arrod {M}illman} (eds.)
  \emph{\bibinfo{booktitle}{{P}roceedings of the 9th {P}ython in {S}cience
  {C}onference}}, \bibinfo{pages}{56 -- 61} (\bibinfo{year}{2010}).

\bibitem{2019AJ....157...98G}
\bibinfo{author}{{Ginsburg}, A.} \emph{et~al.}
\newblock \bibinfo{title}{{astroquery: An Astronomical Web-querying Package in
  Python}}.
\newblock \emph{\bibinfo{journal}{\aj}} \textbf{\bibinfo{volume}{157}},
  \bibinfo{pages}{98} (\bibinfo{year}{2019}).
\newblock \eprint{1901.04520}.

\bibitem{larry_bradley_2023_7946442}
\bibinfo{author}{Bradley, L.} \emph{et~al.}
\newblock \bibinfo{title}{astropy/photutils: 1.8.0} (\bibinfo{year}{2023}).
\newblock \urlprefix\url{https://doi.org/10.5281/zenodo.7946442}.

\bibitem{parrott2020tycho}
\bibinfo{author}{Parrott, D.}
\newblock \bibinfo{title}{Tycho tracker: A new tool to facilitate the discovery
  and recovery of asteroids using synthetic tracking and modern gpu hardware}.
\newblock \emph{\bibinfo{journal}{Journal of the American Association of
  Variable Star Observers (JAAVSO)}} \textbf{\bibinfo{volume}{48}},
  \bibinfo{pages}{262} (\bibinfo{year}{2020}).

\bibitem{Greene2023Nature}
\bibinfo{author}{{Greene}, T.~P.} \emph{et~al.}
\newblock \bibinfo{title}{{Thermal emission from the Earth-sized exoplanet
  TRAPPIST-1 b using JWST}}.
\newblock \emph{\bibinfo{journal}{\nat}} \textbf{\bibinfo{volume}{618}},
  \bibinfo{pages}{39--42} (\bibinfo{year}{2023}).
\newblock \eprint{2303.14849}.

\bibitem{Hoffmann2024}
\bibinfo{author}{Hoffmann, T.} \emph{et~al.}
\newblock \bibinfo{title}{Debiasing astro-photometric observations with
  corrections using statistics (dephocus)}.
\newblock \emph{\bibinfo{journal}{arXiv preprint arXiv:2408.07474}}
  (\bibinfo{year}{2024}).
\newblock \urlprefix\url{https://arxiv.org/abs/2408.07474}.
\newblock \eprint{2408.07474}.

\bibitem{1998Icar..131..291H}
\bibinfo{author}{{Harris}, A.~W.}
\newblock \bibinfo{title}{{A Thermal Model for Near-Earth Asteroids}}.
\newblock \emph{\bibinfo{journal}{\icarus}} \textbf{\bibinfo{volume}{131}},
  \bibinfo{pages}{291--301} (\bibinfo{year}{1998}).

\bibitem{Masiero_2011ApJ...741...68M}
\bibinfo{author}{{Masiero}, J.~R.} \emph{et~al.}
\newblock \bibinfo{title}{{Main Belt Asteroids with WISE/NEOWISE. I.
  Preliminary Albedos and Diameters}}.
\newblock \emph{\bibinfo{journal}{\apj}} \textbf{\bibinfo{volume}{741}},
  \bibinfo{pages}{68} (\bibinfo{year}{2011}).
\newblock \eprint{1109.4096}.

\bibitem{2017A&A...603A..55A}
\bibinfo{author}{{Al{\'\i}-Lagoa}, V.} \& \bibinfo{author}{{Delbo'}, M.}
\newblock \bibinfo{title}{{Sizes and albedos of Mars-crossing asteroids from
  WISE/NEOWISE data}}.
\newblock \emph{\bibinfo{journal}{\aap}} \textbf{\bibinfo{volume}{603}},
  \bibinfo{pages}{A55} (\bibinfo{year}{2017}).
\newblock \eprint{1705.10263}.

\bibitem{2018A&A...612A..85A}
\bibinfo{author}{{Al{\'\i}-Lagoa}, V.}, \bibinfo{author}{{M{\"u}ller}, T.~G.},
  \bibinfo{author}{{Usui}, F.} \& \bibinfo{author}{{Hasegawa}, S.}
\newblock \bibinfo{title}{{The AKARI IRC asteroid flux catalogue: updated
  diameters and albedos}}.
\newblock \emph{\bibinfo{journal}{\aap}} \textbf{\bibinfo{volume}{612}},
  \bibinfo{pages}{A85} (\bibinfo{year}{2018}).
\newblock \eprint{1712.07496}.

\bibitem{2008Icar..193..535W}
\bibinfo{author}{{Wolters}, S.~D.}, \bibinfo{author}{{Green}, S.~F.},
  \bibinfo{author}{{McBride}, N.} \& \bibinfo{author}{{Davies}, J.~K.}
\newblock \bibinfo{title}{{Thermal infrared and optical observations of four
  near-Earth asteroids}}.
\newblock \emph{\bibinfo{journal}{\icarus}} \textbf{\bibinfo{volume}{193}},
  \bibinfo{pages}{535--552} (\bibinfo{year}{2008}).

\bibitem{2011ApJ...743..156M}
\bibinfo{author}{{Mainzer}, A.} \emph{et~al.}
\newblock \bibinfo{title}{{NEOWISE Observations of Near-Earth Objects:
  Preliminary Results}}.
\newblock \emph{\bibinfo{journal}{\apj}} \textbf{\bibinfo{volume}{743}},
  \bibinfo{pages}{156} (\bibinfo{year}{2011}).
\newblock \eprint{1109.6400}.

\bibitem{2012ApJ...744..197G}
\bibinfo{author}{{Grav}, T.} \emph{et~al.}
\newblock \bibinfo{title}{{WISE/NEOWISE Observations of the Hilda Population:
  Preliminary Results}}.
\newblock \emph{\bibinfo{journal}{\apj}} \textbf{\bibinfo{volume}{744}},
  \bibinfo{pages}{197} (\bibinfo{year}{2012}).
\newblock \eprint{1110.0283}.

\bibitem{2012A&A...541A..94V}
\bibinfo{author}{{Vilenius}, E.} \emph{et~al.}
\newblock \bibinfo{title}{{``TNOs are Cool'': A survey of the trans-Neptunian
  region. VI. Herschel/PACS observations and thermal modeling of 19 classical
  Kuiper belt objects}}.
\newblock \emph{\bibinfo{journal}{\aap}} \textbf{\bibinfo{volume}{541}},
  \bibinfo{pages}{A94} (\bibinfo{year}{2012}).
\newblock \eprint{1204.0697}.

\bibitem{2000Icar..148...12P}
\bibinfo{author}{{Pravec}, P.} \& \bibinfo{author}{{Harris}, A.~W.}
\newblock \bibinfo{title}{{Fast and Slow Rotation of Asteroids}}.
\newblock \emph{\bibinfo{journal}{\icarus}} \textbf{\bibinfo{volume}{148}},
  \bibinfo{pages}{12--20} (\bibinfo{year}{2000}).

\bibitem{1989aste.conf..128L}
\bibinfo{author}{{Lebofsky}, L.~A.} \& \bibinfo{author}{{Spencer}, J.~R.}
\newblock \bibinfo{title}{{Radiometry and thermal modeling of asteroids.}}
\newblock In \bibinfo{editor}{{Binzel}, R.~P.}, \bibinfo{editor}{{Gehrels}, T.}
  \& \bibinfo{editor}{{Matthews}, M.~S.} (eds.)
  \emph{\bibinfo{booktitle}{Asteroids II}}, \bibinfo{pages}{128--147}
  (\bibinfo{year}{1989}).

\bibitem{2002aste.book..205H}
\bibinfo{author}{{Harris}, A.~W.} \& \bibinfo{author}{{Lagerros}, J.~S.~V.}
\newblock \bibinfo{title}{{Asteroids in the Thermal Infrared}}.
\newblock In \emph{\bibinfo{booktitle}{Asteroids III}},
  \bibinfo{pages}{205--218} (\bibinfo{year}{2002}).

\bibitem{2014Natur.505..629D}
\bibinfo{author}{{DeMeo}, F.~E.} \& \bibinfo{author}{{Carry}, B.}
\newblock \bibinfo{title}{{Solar System evolution from compositional mapping of
  the asteroid belt}}.
\newblock \emph{\bibinfo{journal}{\nat}} \textbf{\bibinfo{volume}{505}},
  \bibinfo{pages}{629--634} (\bibinfo{year}{2014}).
\newblock \eprint{1408.2787}.

\bibitem{2022PASP..134j5001B}
\bibinfo{author}{{Burdanov}, A.~Y.} \emph{et~al.}
\newblock \bibinfo{title}{{SPECULOOS Northern Observatory: Searching for Red
  Worlds in the Northern Skies}}.
\newblock \emph{\bibinfo{journal}{\pasp}} \textbf{\bibinfo{volume}{134}},
  \bibinfo{pages}{105001} (\bibinfo{year}{2022}).
\newblock \eprint{2209.09112}.

\bibitem{2018SPIE10700E..1ID}
\bibinfo{author}{{Delrez}, L.} \emph{et~al.}
\newblock \bibinfo{title}{{SPECULOOS: a network of robotic telescopes to hunt
  for terrestrial planets around the nearest ultracool dwarfs}}.
\newblock In \bibinfo{editor}{{Marshall}, H.~K.} \&
  \bibinfo{editor}{{Spyromilio}, J.} (eds.)
  \emph{\bibinfo{booktitle}{Ground-based and Airborne Telescopes VII}}, vol.
  \bibinfo{volume}{10700} of \emph{\bibinfo{series}{Society of Photo-Optical
  Instrumentation Engineers (SPIE) Conference Series}},
  \bibinfo{pages}{107001I} (\bibinfo{year}{2018}).
\newblock \eprint{1806.11205}.

\bibitem{Jehin_2011}
\bibinfo{author}{Jehin, E.} \emph{et~al.}
\newblock \bibinfo{title}{Trappist: Transiting planets and planetesimals small
  telescope}.
\newblock \emph{\bibinfo{journal}{The Messenger}}
  \textbf{\bibinfo{volume}{145}}, \bibinfo{pages}{2--6} (\bibinfo{year}{2011}).

\bibitem{2017A&C....18...47M}
\bibinfo{author}{{Mommert}, M.}
\newblock \bibinfo{title}{{PHOTOMETRYPIPELINE: An automated pipeline for
  calibrated photometry}}.
\newblock \emph{\bibinfo{journal}{Astronomy and Computing}}
  \textbf{\bibinfo{volume}{18}}, \bibinfo{pages}{47--53}
  (\bibinfo{year}{2017}).
\newblock \eprint{1702.00834}.

\bibitem{Levine2012}
\bibinfo{author}{{Levine}, S.~E.} \emph{et~al.}
\newblock \bibinfo{title}{{Status and performance of the Discovery Channel
  Telescope during commissioning}}.
\newblock In \bibinfo{editor}{{Stepp}, L.~M.}, \bibinfo{editor}{{Gilmozzi}, R.}
  \& \bibinfo{editor}{{Hall}, H.~J.} (eds.)
  \emph{\bibinfo{booktitle}{Ground-based and Airborne Telescopes IV}}, vol.
  \bibinfo{volume}{8444} of \emph{\bibinfo{series}{Society of Photo-Optical
  Instrumentation Engineers (SPIE) Conference Series}}, \bibinfo{pages}{844419}
  (\bibinfo{year}{2012}).

\bibitem{Christensen_2023LPICo2851.2587C}
\bibinfo{author}{{Christensen}, E.~J.} \emph{et~al.}
\newblock \bibinfo{title}{{Status of the Catalina Sky Survey}}.
\newblock In \emph{\bibinfo{booktitle}{LPI Contributions}}, vol.
  \bibinfo{volume}{2851} of \emph{\bibinfo{series}{LPI Contributions}},
  \bibinfo{pages}{2587} (\bibinfo{year}{2023}).

\end{thebibliography}
\bibliographystyle{naturemag}

\end{document}